\documentclass[a4paper,11pt]{article}
\pdfoutput=1 

\usepackage{jcappub} 
\usepackage{mypkgs}

\title{\boldmath Unified view of scalar and vector dark matter solitons}

\author[a]{Hong-Yi Zhang}

\affiliation[a]{Tsung-Dao Lee Institute \& School of Physics and Astronomy, Shanghai Jiao Tong University, Shanghai 201210, China}

\emailAdd{hongyi18@sjtu.edu.cn}

\abstract{The existence of solitons---stable, long-lived, and localized field configurations---is a generic prediction for ultralight dark matter. These solitons, known by various names such as boson stars, axion stars, oscillons, and Q-balls depending on the context, are typically treated as distinct entities in the literature. This study aims to provide a unified perspective on these solitonic objects for real or complex, scalar or vector dark matter, considering self-interactions and nonminimal gravitational interactions. We demonstrate that these solitons share universal nonrelativistic properties, such as conserved charges, mass-radius relations, stability and profiles. Without accounting for alternative interactions or relativistic effects, distinguishing between real and complex scalar dark matter is challenging. However, self-interactions differentiate real and complex vector dark matter due to their different dependencies on the macroscopic spin density of dark matter waves. Furthermore, gradient-dependent nonminimal gravitational interactions impose an upper bound on soliton amplitudes, influencing their mass distribution and phenomenology in the present-day universe.}

\begin{document}
\maketitle
\flushbottom

\section{Introduction}
Ultralight bosons as dark matter were first proposed to solve the small-scale problems associated with cold dark matter \cite{Hu:2000ke}. The long de Broglie wavelength of these particles suppresses the formation of small-scale structures, offering a natural solution to the cusp-core problem, which refers to the mismatch between the cuspy density profile predicted by N-body simulations and the flatter ones observed in galactic centers \cite{Ferreira:2020fam}. With mass much less than $40\rm{eV}$, ultralight dark matter can be described by classical field theory, and its wave mechanics is governed by the Schroedinger equation coupled to gravity through Poisson's equation \cite{Hui:2021tkt, Zhang:2023ktk, Matos:2023usa}. Simulating such a wave system, it has been shown that the density profile of each dark matter halo can be well fitted by the Navarro-Frenk-White profile \cite{Navarro:1995iw} at large radii, while the core is described by a soliton \cite{Schive:2014dra, Levkov:2018kau, Veltmaat:2018dfz, Mocz:2019pyf, May:2021wwp, Gorghetto:2022sue, Chen:2020cef, Amin:2022pzv, Chen:2023bqy, Gorghetto:2024vnp}.

Here, we refer to solitons as stable, long-lived, and localized field configurations whose boundary conditions at infinity are the same as those for the physical vacuum state.\footnote{More generally, there are topological solitons whose boundary conditions at infinity are topologically different from those of the physical vacuum state.} In the literature, they sometimes appear with different names depending on the contexts. If the mass of fields is not in the fuzzy regime, i.e., with mass $m\gg 10^{-20}\rm{eV}$, solitons could be cold stellar configurations called boson, Bose, or soliton stars \cite{Lee:1991ax, Breit:1983nr, Seidel:1991zh}.\footnote{In early examples, complex scalar solitons were called Klein-Gordon geons \cite{Kaup:1968, ruffini1969systems}.} Solitons constituting axions or axionlike particles are also called dense or dilute axion stars \cite{Braaten:2015eeu, Schiappacasse:2017ham, Visinelli:2017ooc, Hertzberg:2020dbk, Chavanis:2017loo}, depending on whether the self-interactions (SIs) of the fields are important or not. Using the Klein-Gordon and Einstein equations for scalar fields, solitons are sometimes called oscillatons \cite{Seidel:1993zk, Alcubierre:2003sx}. In cases where attractive SIs dominate over gravity, they can be referred to as Q-balls \cite{Coleman:1985ki, Kusenko:1997ad} if they are complex-valued or oscillons/I-balls/quasi-breathers \cite{Copeland:1995fq, Kasuya:2002zs, Saffin:2006yk} if they are real-valued.\footnote{In 1-dimensional space, there could exist exact periodic solutions for real scalar field equations called breathers, e.g., breathers in the sine-Gordon equation. In 3-dimensional space, there are only approximate periodic solutions which slowly emit classical radiation \cite{Fodor:2008es, Fodor:2008du, Zhang:2020bec, Zhang:2020ntm}.} For massive vector fields, their corresponding solitons are sometimes called Proca stars \cite{Brito:2015pxa}, Proca Q-balls \cite{Loginov:2015rya}, vector solitons \cite{Adshead:2021kvl, Jain:2021pnk}, or vector oscillons \cite{Zhang:2021xxa}. In the literature, these solitonic objects are typically regarded as different entities and their phenomenological implications are often discussed in an uncorrelated manner. For reviews, see \cite{Liebling:2012fv, Visinelli:2021uve}.

Considering their diverse and distinct applications, in this work, I aim to provide a unified understanding of these solitonic objects in the nonrelativistic regime, e.g., dark matter solitons. A clear connection between these solitons will be established by examining field equations governing the nonrelativistic dynamics of real or complex, scalar or vector dark matter. This extends the previous result that oscillons have an approximate conserved particle number, e.g., see \cite{Mukaida:2014oza, Mukaida:2016hwd, Zhang:2021xxa}. Notably, we will see that real vector dark matter provides the most general scenario, in the sense that its field equations could reduce to the other three cases through a few simple manipulations of variables. Therefore, we will use it as an example to explore the universal properties of dark matter solitons, such as conserved quantities, mass-radius relation, stability, and profiles. 

In addition to the SIs of dark matter, I will also take into consideration its nonminimal gravitational interactions (NGIs). They could arise from quantum corrections and are essential for the renormalization of field theories in curved spacetime \cite{Birrell:1982ix, Weinberg:1995mt, callan1970new, freedman1974energy, FREEDMAN1974354}. Their phenomenological implications have been studied in various contexts, such as dark matter \cite{Ivanov:2019iec, Ji:2021rrn, Sankharva:2021spi, Barman:2021qds, Zhang:2023fhs}, inflation \cite{Starobinsky:1979ty, Turner:1987bw, Ford:1989me, Faraoni:2000wk, Golovnev:2008cf, Golovnev:2008hv, Golovnev:2009ks, Golovnev:2009rm}, and modified gravity \cite{Moffat:2005si, Brownstein:2005zz, Tasinato:2014eka, Heisenberg:2014rta, DeFelice:2016yws, deFelice:2017paw}.

For the rest of the paper, we derive the nonrelativistic effective field theory for real or complex, scalar or vector dark matter in section \ref{sec:eft}. The symmetries preserved by the nonrelativistic effective actions and the associated conserved charges are discussed in appendix \ref{sec:conserved_charges}. Soliton equations are derived, and the impacts of SIs and NGIs are qualitatively discussed in section \ref{sec:vec_soliton}. In section \ref{sec:mr_relation}, we explore the stability and mass-radius relation both analytically and numerically. Soliton solutions are numerically calculated using the Mathematica package ``DMSolitonFinder'', whose details are provided in appendix \ref{sec:mathematica}. Throughout the paper, I use the natural units in high-energy physics with $c\equiv \hbar\equiv 1$. The reduced Planck mass is defined as $\MP\equiv (8\pi G)^{-1/2}$. Repeated indices are summed unless otherwise stated.

\section{Nonrelativistic effective field theory}
\label{sec:eft}
It is relatively simple and insightful to study ultralight dark matter and its solitons using a nonrelativistic effective field theory, which separates the dominant nonrelativistic dynamics from small relativistic effects. In this section, I will start with the relativistic actions for real or complex, scalar or vector dark matter fields, including SIs and NGIs, and derive their effective actions and field equations in the nonrelativistic regime. We will see that the model of real vector dark matter offers the most general scenario, capable of reducing to the other three cases through a few simple manipulations of variables. Based on the nonrelativistic actions, conserved charges such as particle number and angular momentum will be clarified.

A full action $S$ may be decomposed into a minimal gravity part and a matter part, $S = S_\rm{G} + S_\rm{M}$, where the former is given by
\begin{align}
	S_\rm{G} = \int d^4x \sqrt{-g} ~ \frac{\MP^2}{2} R ~,
\end{align}
with $R$ being the Ricci scalar. In the nonrelativistic regime, the metric can be written as \cite{Baumann:2022mni}
\begin{align}
	ds^2 = -(1+2A) dt^2 + a^2 (1-2A) \delta_{ij} dx^i dx^j ~,
\end{align}
where $A(t,\b x)$ is a scalar perturbation and $a(t)$ is the cosmological scale factor related to the redshift $z$ by $a=(1+z)^{-1}$. In an expanding universe, it is convenient to work with a comoving scalar perturbation $\Phi \equiv aA$. Since dark matter fields are dominated by oscillations with a frequency $\simeq m$, it is useful to perform a comoving and nonrelativistic expansion for dark matter fields,
\begin{align}
	\label{nr_expansion_re_sca}
	\text{Real scalars:}\quad & \f(t,\b x) = \frac{1}{\sqrt{2ma^3}} \[ e^{-imt} \psi(t,\b x) + e^{imt} \psi^*(t,\b x) \] ~,\\
	\label{nr_expansion_co_sca}
	\text{Complex scalars:}\quad & \f(t,\b x) = \frac{1}{\sqrt{2ma^3}} e^{-imt} \psi(t,\b x) ~,\\
	\label{nr_expansion_re_vec}
	\text{Real vectors:}\quad & X_\mu(t,\bx) = \frac{1}{\sqrt{2ma}} \[ e^{-imt} \psi_\mu(t,\bx) + e^{imt} \psi_\mu^*(t,\bx) \] ~,\\
	\label{nr_expansion_co_vec}
	\text{Complex vectors:}\quad & X_\mu(t,\bx) = \frac{1}{\sqrt{2ma}} e^{-imt} \psi_\mu(t,\bx) ~,
\end{align}
where $m$ is the field mass, and $\psi$ and $\psi_\mu$ are complex comoving fields that slowly vary in time. Here I have denoted real and complex fields with the same letters, but the meaning of the notations should be clear based on the context. To ensure that the redefinition for real scalars preserves the number of propagating degrees of freedom, one can employ a constraint that keeps the field equation of $\psi$ first-order in time derivatives, $e^{-imt} \dot\psi + e^{imt} \dot\psi^* = 0$ \cite{Salehian:2020bon}, and a similar constraint can be used for real vectors. In what follows, I will give a brief review of a systematic approach for power counting based on the comoving and nonrelativistic fields and then derive the leading-order low-energy effective actions and field equations for dark matter.

\subsection{Power counting}
As we take the nonrelativistic limit of a relativistic theory, several small parameters or operators appear, allowing us to organize different terms that arise in the effective theory. Taking real scalar fields as an example, we can follow the prescription in \cite{Salehian:2021khb} and identify the following dimensionless parameters or operators,
\begin{align}
	\label{small_parameter1}
	\epsilon_H \equiv \frac{H}{m} \sep
	\epsilon_t \equiv \abs{ \frac{\pd_t Q}{m Q} } \sep
	\epsilon_k \equiv \abs{ \frac{\nabla^2 Q}{a^2 m^2 Q} } \sep
	\epsilon_\psi \equiv \frac{|\phi|}{\MP} \sim \frac{|\psi|}{a^{3/2} m^{1/2} \MP} \sep
	\epsilon_\Phi \equiv |\Phi| ~,
\end{align}
where $H$ is the Hubble parameter and $Q$ can be any of the comoving and slowly varying variables, including $a, H, \psi, \Phi$. These parameters must be less than unity, since in the nonrelativistic regime a mass term is the dominant contribution to the time evolution of the original field and all other effects are suppressed. If the relativistic action of dark matter includes a quartic SI term, $\lambda\phi^4$, and a NGI term, $\xi\phi^2R$, we must have two more small parameters
\begin{align}
	\label{small_parameter2}
	\epsilon_\lambda \equiv \frac{|\lambda| \phi^2}{m^2} \sep
	\epsilon_\xi \equiv \abs{\frac{\xi R}{m^2}} ~.
\end{align}
Since the comoving mass density $\rho\sim a^3 m^2\phi^2$ and $R\approx 3H^2 + 2 a^{-3}\nabla^2\Phi \sim a^{-3} \MP^{-2} \rho$, the requirement of $\epsilon_\lambda, \epsilon_\xi$ being small at the radiation-matter equality sets an upper limit on $|\lambda|$ and $|\xi|$,
\begin{align}
	\label{uplimit_lambda}
	|\lambda| &\ll \frac{m^4}{(1+z)^3 \rho} = 3\times 10^{-72} \( \frac{m}{10^{-18} ~\rm{eV}} \)^4 \( \frac{10^{-6} ~\GeV/\rm{cm}^3}{\rho} \) \( \frac{3400}{1+z} \)^3 ~,\\
	\label{uplimit_xi}
	|\xi| &\ll \frac{m^2\MP^2}{(1+z)^3 \rho} = 2\times 10^{19} \( \frac{m}{10^{-18} ~\rm{eV}} \)^2 \( \frac{10^{-6} ~\GeV/\rm{cm}^3}{\rho} \) \( \frac{3400}{1+z} \)^3 ~.
\end{align}
In general, the small parameters are not independent of each other and we do not know a priori the relative magnitudes between them, thus a reliable power counting strategy would require the system of equations to be expanded up to a homogeneous order of all small parameters. For notation convenience, let us denote all small parameters collectively by $\epsilon=\{ \epsilon_H, \epsilon_t, \epsilon_k, \epsilon_\psi, \epsilon_\Phi, \epsilon_\lambda, \epsilon_\xi \}$. The impact of relativistic modes on the nonrelativistic modes is then regulated by the largest value of $\epsilon$. 

Due to the oscillating factors in the equations of motion, the dynamics of slowly varying quantities are affected by rapidly oscillating terms, whose frequencies are typically integer multiples of the field mass $m$. Generally speaking, a nonrelativistic variable $Q$ can be expanded into
\begin{align}
	\label{fourier_expansion}
	Q = \sum_{\nu=-\infty}^\infty Q_\nu e^{i\nu m t} ~,
\end{align}
where $Q_\nu$ are modes that vary slowly in time and typically $Q_0$ dominates over nonzero modes. For the purpose of this work, we will only be interested in the zero modes and thus need to integrate out the rest of the modes. To systematically solve for the nonzero modes using power counting, we expand them into a power series in $\epsilon$,
\begin{align}
	Q_\nu = \sum_{n=1}^{\infty} Q_\nu^{(n)} \quad (\nu\neq 0) ~,
\end{align}
where the superscript $(n)$ indicates the term's order of magnitude relative to the zero mode, i.e., $|Q_\nu^{(n)}/Q_0| \sim \cal O(\epsilon^n)$. By plugging these expansions for all comoving and nonrelativistic variables into their field equations, we can solve the nonzero modes perturbatively and then substitute the solutions back into the equations for the zero modes. The detailed procedure is outlined in \cite{Salehian:2021khb}. Up to the leading-order terms containing $\Phi$, the gravity action becomes
\begin{align}
	\label{nr_action_gravity}
	S_\rm{G,NR} = \int d^4x ~ \MP^2 a (-3\dot a^2 + a^{-2} \Phi \nabla^2\Phi - 6\ddot a\Phi) ~.
\end{align}
For clarity in notation, here I have omitted the subscript ``$_0$'' on the zeros modes defined in \eqref{fourier_expansion} and will apply the same approach to other nonrelativistic fields. Let us now find nonrelativistic matter actions using similar techniques.

\subsection{Scalar dark matter}
If the dominant component of dark matter is real scalar particles such as axions, it may be described by the action
\begin{align}
	\label{action_re_sca}
	S_\rm{M} = \int d^4x \sqrt{-g} \[ - \frac{1}{2}\pd_\mu\f \pd^\mu\f - \frac{m^2}{2}\f^2 - \frac{\lambda}{4!}\f^4 - \frac{\xi}{2} R \f^2 \] ~,
\end{align}
where a quartic SI term characterized by $\lambda$ and a NGI term characterized by $\xi$ are included. By plugging the field redefinition \eqref{nr_expansion_re_sca} into the matter action and integrating out fast oscillating modes, we can obtain an effective action for the slow modes. To the leading order in $\epsilon$, it becomes
\small
\begin{align}
	\label{nr_action_re_sca}
	S_\rm{M,NR} = \int d^4x \bigg [ i\dot\psi \psi^* + \frac{1}{2a^2 m}(\nabla^2\psi) \psi^* - \frac{m}{a}\Phi |\psi|^2 - \frac{\lambda}{16a^3 m^2} |\psi|^4 - \frac{\xi}{m} \(\frac{\nabla^2\Phi}{a^3} + 3 H^2 + 3\frac{\ddot a}{a} \) |\psi|^2 \bigg ] ~.
\end{align}
\normalsize
In deriving this leading-order action in $\epsilon$, the gauge constraint $e^{-imt} \dot\psi + e^{imt} \dot\psi^*=0$ is not directly used since we treat $\epsilon_t$ as a small quantity and integrate out fast oscillating modes. But in higher orders, it would be crucial in removing the redundancy in the definition of $\psi$ of the form, $\psi\rightarrow \psi + i e^{imt} \eta$ with $\eta$ being any real function of spacetime, which leaves $\phi$ invariant \cite{Salehian:2020bon}. The $\lambda$- and $\xi$-dependent terms are meaningful only if their magnitudes are larger than relativistic corrections, equivalently $\epsilon_\lambda \gg \epsilon_\psi^2$ and $\epsilon_\xi \gg \epsilon_\psi^2$, which give
\begin{align}
	\label{lowlimit_lambda_xi}
	|\lambda| \gg \frac{m^2}{\MP^2} = 2\times 10^{-91} \( \frac{m}{10^{-18} ~\rm{eV}} \)^2 \sep
	|\xi| \gg 1 ~.
\end{align}
Field equations can be obtained by varying the nonrelativistic gravity action \eqref{nr_action_gravity} and matter action \eqref{nr_action_re_sca} with respect to $a$, $\psi^*$, and $\Phi$,
\begin{align}
	\label{eom_re_sca1}
	i\pd_t\psi = -\frac{\nabla^2}{2ma^2}\psi + \frac{m}{a}\Phi \psi + \frac{\lambda}{8a^3m^2} |\psi|^2\psi + \frac{\xi}{2a^3 m\MP^2} \rho_\xi \psi ~,\\
	\label{eom_re_sca2}
	\nabla^2\Phi = \frac{1}{2\MP^2} (\rho_\xi - \bar\rho_\xi) \sep
	3\MP^2 H^2 = \frac{\bar\rho_\xi}{a^3} ~,
\end{align}
where the overline stands for spatial averaging and the comoving mass density $\rho_\xi$ is
\begin{align}
	\rho_\xi \equiv \rho + \frac{\xi}{a^2m^2} \nabla^2\rho \sep
	\rho \equiv m|\psi|^2 ~.
\end{align}
It turns out that in the nonrelativistic regime, the NGI becomes a gradient-dependent SI of fields as well as a correction to the Poisson equation for the gravitational potential. One may decompose $\Phi$ into the Newtonian potential $\Phi_\rm{N}$ and the $\xi$-dependent part $\Phi_\xi$, where $\nabla^2\Phi_\rm{N} = (\rho - \bar\rho)/(2\MP^2)$ and $\Phi_\xi = \xi\rho/(2a^2m^2\MP^2)$. To be consistent with solar system tests, we should demand $\Phi_\xi \ll \Phi_\rm{N}$ and this gives another upper limit of $|\xi|$ different from \eqref{uplimit_xi},
\begin{align}
	\label{uplimit_xi1}
	|\xi| \ll \frac{2 m^2 \MP^2 \Phi_\rm{N}}{\rho} = 4\times 10^{18} \( \frac{m}{10^{-18} ~\rm{eV}} \)^2 \( \frac{0.4 ~\GeV/\rm{cm}^3}{\rho} \) \( \frac{\Phi_\rm{N}}{10^{-6}} \) ~.
\end{align}
Additionally, to retain the success of cold dark matter in explaining matter power spectrum on large scales (for Fourier modes with $k\lesssim 10 h\rm{Mpc^{-1}}$ \cite{Irsic:2017yje, Bechtol:2022koa}), the NGI coupling should satisfy \cite{Zhang:2023fhs}
\begin{align}
	\label{uplimit_xi2}
	|\xi|\ll 10^{14} \( \frac{m}{10^{-18} ~\rm{eV}} \)^{2} ~.
\end{align}
The equations \eqref{uplimit_lambda}, \eqref{uplimit_xi}, \eqref{lowlimit_lambda_xi}, \eqref{uplimit_xi1}, and \eqref{uplimit_xi2} collectively establish the validity conditions of our nonrelativistic effective field theory.

If the dominant component of dark matter is complex scalar particles with an action
\begin{align}
	\label{nr_action_co_sca}
	S_\rm{M} = \int d^4x \sqrt{-g} \[ - \pd_\mu\f^\da \pd^\mu\f - m^2\f^\da\f - \frac{\lambda}{4}(\f^\da \f)^2 - \xi R \f^\da\f \] ~,
\end{align}
one can substitute the nonrelativistic field expansion \eqref{nr_expansion_co_sca} into the action and derive a nonrelativistic action for slow modes, which is identical to \eqref{nr_action_re_sca}. In this context, it is very challenging to discern the nature of scalar dark matter without accounting for alternative interactions or relativistic effects.

\subsection{Vector dark matter}
\label{sec:vdm}
If dark matter can be described by real vector fields, the action may be written as
\begin{align}
	\label{action_re_vec}
	S_\rm{M} = \int d^4x \sqrt{-g} \[ -\frac{1}{4} X_{\mu\nu} X^{\mu\nu} - \frac{m^2}{2} X_\mu X^\mu - \frac{\lambda}{4!} (X_\mu X^\mu)^2 - \frac{\xi_1}{2} R X_\mu X^\mu - \frac{\xi_2}{2} R^{\mu\nu} X_\mu X_\nu  \] ~,
\end{align}
where $X_{\mu\nu} = \pd_\mu X_\nu - \pd_\nu X_\mu$, and a quartic SI term characterized by $\lambda$ and NGI terms characterized by $\xi_1$ and $\xi_2$ are included. By plugging the field redefinition \eqref{nr_expansion_re_vec} into the matter action and keeping terms that are leading-order in $\epsilon$, we obtain an effective action for the slow modes
\begin{align}
	\label{nr_action_re_vec}
	\nonumber
	S_\rm{M,NR} = &\int d^4x ~\bigg\{ \frac{a^2 m}{2} \abs{\psi_0 - \frac{i}{a^2 m}\nabla\cdot\b\psi}^2 + i \dot{\b \psi}\cdot\b \psi^* + \frac{1}{2a^2 m} (\nabla^2\b\psi)\cdot\b\psi^* - \frac{m}{a}\Phi \abs{\b\psi}^2 \\
	& - \frac{\lambda}{16a^3m^2} \( \abs{\b\psi}^4 - \frac{\b{\cal S}^2}{3} \) - \[ \xi \frac{\nabla^2\Phi}{a^3} + \xi_1 \(3H^2 + 3\frac{\ddot a}{a}\) + \frac{\xi_2}{2} \(2H^2 + \frac{\ddot a}{a} \) \] \frac{|\b\psi|^2}{m} \bigg\} ~,
\end{align}
where $\xi\equiv \xi_1 + \xi_2/2$, the comoving spin density is defined as $\cal S_i\equiv i\ve_{ijk} \psi_j \psi_k^*$ with $\ve_{ijk}$ being the Levi-Civita symbol, and $\b{\cal S}^2 = |\b\psi|^4 - |\b\psi\cdot\b\psi|^2$. Since we focus on nonrelativistic dynamics, this model avoids problems such as the violation of perturbative unitarity \cite{Schwartz:2014sze}, the ghost instabilities of longitudinal modes \cite{Himmetoglu:2009qi, Himmetoglu:2008zp, Himmetoglu:2008hx, Esposito-Farese:2009wbc, Graham:2015rva}, the singularity problem \cite{Mou:2022hqb, Clough:2022ygm, Coates:2022qia}, and the runaway production of high-momentum modes \cite{Capanelli:2024pzd}. By varying the nonrelativistic gravity action \eqref{nr_action_gravity} and matter action \eqref{nr_action_re_vec} with respect to $a$, $\psi_i^*$, and $\Phi$, we obtain the field equations
\begin{align}
	\label{eom_re_vec1}
	i\pd_t \psi_i = - \frac{\nabla^2}{2ma^2} \psi_i + \frac{m}{a}\Phi \psi_i + \frac{\lambda}{8a^3m^2}\( \abs{\b\psi}^2 \psi_i + \frac{i}{3} \ve_{ijk} \psi_j \cal S_k \) + \frac{\xi}{2a^3 m \MP^2} \rho_\xi \psi_i ~,\\
	\label{eom_re_vec2}
	\nabla^2\Phi = \frac{1}{2\MP^2} (\rho_\xi - \bar\rho_\xi) \sep 3\MP^2H^2 = \frac{\bar\rho_\xi}{a^3} ~,
\end{align}
where the comoving mass density is
\begin{align}
	\rho_\xi \equiv \rho + \frac{\xi}{a^2m^2} \nabla^2\rho \sep
	\rho \equiv m|\b \psi|^2 ~.
\end{align}
We can see that the vector field equations \eqref{eom_re_vec1} and \eqref{eom_re_vec2} could reduce to those for scalars \eqref{eom_re_sca1} and \eqref{eom_re_sca2} through the replacement $\psi_i\rightarrow \psi$ and $\xi_2\rightarrow 0$.

Now let us examine the case where dark matter is comprised of complex vector particles, whose action is given by
\begin{align}
	\label{action_co_vec}
	S_\rm{M} = \int d^4x \sqrt{-g} \[ -\frac{1}{2} X_{\mu\nu}^* X^{\mu\nu} - m^2 X_\mu^* X^\mu - \frac{\lambda}{4} (X_\mu^* X^\mu)^2 - \xi_1 R X_\mu^* X^\mu - \xi_2 R^{\mu\nu} X_\mu^* X_\nu  \] ~.
\end{align}
As we will see, this action has a $U(1)$ symmetry for each component of the field in orthonormal polarization bases (e.g., the longitudinal and transverse modes) in the nonrelativistic limit.\footnote{Because of this reason, the dimension-4 operator $|X_\mu X^\mu|^2$ is not explicitly considered. The inclusion of such a term would yield additional contributions to the effective quartic coupling $\til\lambda_\sigma$ in the soliton profile equation \eqref{soliton_profile_eq_v1}, as the nonrelativistic action \eqref{nr_action_re_vec} has included all types of quartic self-interactions invariant under $SO(3)$. This leaves the main conclusion of this paper---that solitons share universal nonrelativistic properties---unchanged. The phenomenological consequence of the term will be left for future work.} By substituting the nonrelativistic field expansion \eqref{nr_expansion_co_vec} into the action, we can derive a nonrelativistic action for slow modes, which corresponds to \eqref{nr_action_re_vec} but with the spin density term $\b{\cal S}^2$ set to zero. The nonrelativistic action \eqref{nr_action_re_vec} now is manifestly invariant under $U(1)$ symmetry for the longitudinal or transverse mode of the vector field.

In summary, under the current setup, the model of real vector dark matter provides the most general scenario, capable of being reduced to the other three cases -- where dark matter consists of real scalars, complex scalars, or complex vectors -- through a few simple manipulations of variables. Thus in the discussions that follow, we may consider real vector solitons as a general model of dark matter solitons. The symmetries preserved by the nonrelativistic action \eqref{nr_action_re_vec} and the associated conserved charges are discussed in appendix \ref{sec:conserved_charges}.

For both analytical and numerical studies, it is judicious to work with dimensionless quantities to reduce the number of variables. Let us nondimensionalize the real vector equations by making the replacement
\begin{align}
	\label{nondimensionalize}
	t \rightarrow \frac{\til t }{v^2 m}\sep
	\b x \rightarrow \frac{\til{\b x}}{v m} \sep
	\Phi \rightarrow v^2 \til\Phi \sep
	\psi_i \rightarrow v^2 \sqrt{m} F \til \psi_i \sep
	\lambda \rightarrow \frac{m^2 \til \lambda}{v^2 F^2} \sep
	\xi \rightarrow \frac{\til\xi}{v^2} ~,
\end{align}
where $v$ is an arbitrary positive number and $F$ is an arbitrary mass scale. Quantities with a tilde on top are dimensionless and can be interpreted as the corresponding dimensional quantities in specific units. For example, $\til{\b x}$ can be regarded as $\b x$ in units of $(mv)^{-1}$. Then the equations \eqref{eom_re_vec1} and \eqref{eom_re_vec2} become
\begin{align}
	\label{eom_re_vec_tilde1}
	i \pd_{\til t} \til\psi_i = - \frac{\til\nabla^2}{2a^2} \til\psi_i + \frac{\til\Phi}{a} \til\psi_i + \frac{\til \lambda}{8a^3} \(|\til{\b \psi}|^2 \til\psi_i + \frac{i}{3} \ve_{ijk} \til\psi_j \til{\cal S}_k\) + \til\xi \til\rho_\xi \til\psi_i ~,\\
	\label{eom_re_vec_tilde2}
	\til\nabla^2 \til\Phi = \til\rho_\xi - \bar{\til\rho}_\xi \sep
	\frac{3}{2} \til H^2 = \frac{\bar{\til\rho}_\xi}{a^3} ~,
\end{align}
where $\til{\cal S}_i = i\ve_{ijk} \til\psi_j \til\psi_k^*$, $\til\rho_\xi = \til\rho + \til\xi\til\nabla^2\til\rho$, $\til\rho = |\til{\b\psi}|^2$, and $F^2$ is taken to be $2\MP^2$. These equations are independent of $v$, implying a scaling symmetry that can be used to set the value of either $|\til\lambda|$ or $|\til\xi|$ (but not both) to unity.

\section{Real vector solitons as a general model}
\label{sec:vec_soliton}
Solitons are states at the local extremum of energy for a fixed particle number \cite{Lee:1991ax}. For vector solitons, this requires the field to have the time dependence $\til\psi_i \propto e^{i\til\mu \til t}$ \cite{Lee:1991ax, Zhang:2023ktk}. In the rest frame (i.e., the nonrelativistic limit) of a massive wave, we may decompose it into different polarization or spin states,\footnote{More generally, the polarization vector depends on the momentum of waves. In quantum field theory, a vector field with the polarization vector $e_i(\sigma)$ would excite particles with a spin $\sigma$ in the $z$-direction \cite{Weinberg:1995mt}.}
\begin{align}
	\label{soliton_ansatz}
	\til\psi_i(\til t, \til\bx, \sigma) = e_i(\sigma) \til f(\til\bx, \sigma) e^{i\til\mu \til t} ~,
\end{align}
where $e_i(\sigma)$ stands for orthonormal base polarization vectors and $\sigma=0,\pm 1$ characterizes the polarization with respect to the $z$-direction. The base vectors $e_i(\sigma)$ can be chosen as $e_i(0) = (0,0,1)$ and $e_i(\pm 1) = (1,\pm i, 0) /\sqrt{2}$, and solitons with $\sigma=0$ or $\sigma=\pm 1$ polarization are referred to as linearly or circularly polarized. In these solitons, the spatial part of the original field $X_i$ at each location oscillates along a fixed direction or moves in a circle; for this reason, they may also be called directional or spinning solitons. For scalar solitons, their fields have the same form as \eqref{soliton_ansatz} but with $\sigma\rightarrow 0$ and $e_i\rightarrow 1$.

Since dark matter solitons are localized objects within dark matter halos, we ignore the expansion of the universe for the remainder of the discussions. In this section, let us focus on understanding the general properties of solitons without doing numerical calculations. I will first present a brief review of basic equations and properties for extremally polarized solitons corresponding to \eqref{soliton_ansatz}, then discuss in what cases there could be partially polarized solitons. More details can be found in references \cite{Jain:2021pnk, Zhang:2021xxa}. After that, I will make analytical arguments regarding the impacts of SIs and NGIs. Numerical calculations for soliton profiles and mass-radius relation will be carried out in section \ref{sec:mr_relation}.

\subsection{Extremally polarized vector solitons}
In the presence of SIs, the ground state of real vector solitons among all configurations is extremally polarized and takes the form specified in equation \eqref{soliton_ansatz} \cite{Zhang:2021xxa}.\footnote{This statement does not hold for complex vector solitons since we have seen in section \ref{sec:vdm} that the spin density does not directly differentiate the energy between solitons.} Assuming that the soliton profile $\til f$ is radially symmetric, it is straightforward to obtain the profile equation by plugging \eqref{soliton_ansatz} into the field equations \eqref{eom_re_vec_tilde1} and \eqref{eom_re_vec_tilde2},
\begin{align}
\label{soliton_profile_eq_v1}
-\frac{1}{2}\til\nabla^2 \til f + \til\Psi \til f + \frac{\til\lambda_\sigma}{8} \til\rho \til f + \til\xi \til\rho_\xi \til f = 0 \sep
\til\nabla^2 \til \Psi = \til\rho_\xi  ~,
\end{align}
where $\til\Psi \equiv \til\mu + \til\Phi$, $\til\lambda_\sigma\equiv (1-\sigma^2/3)\til\lambda$, and $\til\rho = \til f^2$. Thus a nonzero polarization $\sigma$ effectively diminishes the SI coupling for real vector solitons. It is also illuminating to rewrite the equations as
\begin{align}
\label{soliton_profile_eq_v2}
-\frac{1}{2}\til\nabla^2 \til f + \til\Psi_\rm{N} \til f + \frac{\til\gamma}{8} \til\rho \til f + \til\xi^2 (\til\nabla^2\til\rho) \til f = 0 \sep
\til\nabla^2 \til \Psi_\rm{N} = \til\rho  ~,
\end{align}
where $\til\gamma \equiv \til\lambda_\sigma + 16\til\xi$ and I have decomposed $\Psi$ into the Newtonian part $\til\Psi_\rm{N}$ and the $\xi$-dependent part $\til\Psi_\xi$, where $\til\Psi_\xi = \til\xi\til\rho$.\footnote{Without tilde, the definition becomes $\gamma\equiv \lambda_\sigma + 8(m^2/\MP^2) \xi$ with $\gamma\rightarrow [m^2/v^2 F^2] \til\gamma$.}  The interaction terms in \eqref{soliton_profile_eq_v2} are grouped based on their physical properties. For example, the $\til\Psi_\rm{N}$ term describes the standard Newtonian gravity in the absence of NGIs, and the $\til\gamma$ and $\til\xi^2$ terms characterize the gradient-independent and -dependent SIs. In comparison, the interaction terms in \eqref{soliton_profile_eq_v1} are grouped in terms of their physical origins, i.e., $\til\lambda$ and $\til\xi$ come from different ultraviolet physics. The profile equations \eqref{soliton_profile_eq_v1} and \eqref{soliton_profile_eq_v2} would reduce to those for scalars or complex vectors if $\til\lambda_\sigma \rightarrow \til\lambda$.

Since soliton profiles must be smooth at the center, implying $\pd_{\til r}\til f|_{\til r=0}= \pd_{\til r}\til \Psi|_{\til r=0} = 0$, numerical soliton solutions can be found by solving the profile equations \eqref{soliton_profile_eq_v1} or \eqref{soliton_profile_eq_v2} with different values of $\til f|_{\til r=0}$ or $\til \Psi|_{\til r=0}$ until the solutions match the boundary condition $\til f\rightarrow 0$ at $\til r\rightarrow \infty$. Once a set of soliton profiles is obtained, its oscillating frequency $\til\mu$ can be inferred by noting $\til r\til\Phi \rightarrow \text{const}$ for $\til r\rightarrow \infty$.

With a set of soliton solutions, we can calculate its dimensionless mass and particle number using
\begin{align}
	\label{soliton_mass}
	\til M_s = \int d^3\til x ~ \til\rho_\xi \sep
	\til N_s = \int d^3\til x ~ \til n ~,
\end{align}
where the number density is defined as $\til n \equiv \til f^2$. We may define the radius of a soliton $\til R_s$ to be the one enclosing $95\%$ of its mass. The impact of a nonzero $\til\xi$ is not only on solitons' profiles but also on their radii---solitons with identical profiles and mass could have different radius due to the existence of $\til\xi$. For extremally polarized solitons, their spin components in different directions \eqref{soliton_ansatz} are
\begin{align}
	\til S_{s,1} = 0 \sep
	\til S_{s,2} = 0 \sep
	\til S_{s,3} = \sigma \til N_s ~,
\end{align}
where $\til S_{s,i} = \int d^3\til x ~\til{\cal S}_i$. Therefore, we see that linearly and circularly polarized solitons are constituted by particles with a spin $\sigma$ in the $z$-direction.

\subsection{Impacts of polarization-dependent self-interactions}
The polarization dependence of the SI coupling $\lambda_\sigma$ results in two significant consequences for the properties of solitons: It prevents the superposition of extremally polarized soliton fields and differentiates their energy, defined as
\begin{align}
	\label{soliton_energy}
	\til E_s = \int d^3\til x ~ \[ \frac{1}{2} \til\pd_j\til\psi_i \til\pd_j\til\psi_i^*  + \frac{1}{2}\til\Phi\til\rho_\xi + \frac{1}{16}\til\lambda \(1- \frac{ \til{\b{\cal S}}^2 }{3|\b\psi|^4} \) |\til{\b\psi}|^4 \] ~,
\end{align}
at a given particle number. I will now elaborate on these statements.

If the polarization $\sigma$ does not explicitly appear in the soliton profile equations \eqref{soliton_profile_eq_v1} and \eqref{soliton_profile_eq_v2}, as is the case for complex vector solitons, then extremally polarized solitons will have identical profiles for a given frequency $\til\mu$. Under these conditions, it is possible to have partially polarized solitons that share the same profile but with a polarization vector that represents a superposition of the base vectors
\begin{align}
	\label{polarization_superposition}
	e_i(s, \sigma) \equiv \sum_{\sigma' = 0,\pm 1} a_{\sigma'} e_i(\sigma') ~,
\end{align}
where the polarization numbers $s \equiv |\til{\b{\cal S}}|/\til n $ and $\sigma = \til{\cal S}_{3}/\til n$ represent the total spin and its $z$-component for a particle in the soliton, thus $s \in [0,1]$ and $\sigma \in [-1,1]$, and $a_{\sigma'}$ are arbitrary numbers normalized by $\sum_{\sigma'} |a_{\sigma'}|^2=1$. It is straightforward to show
\begin{align}
	s^2 = (|a_{-1}|^2 - |a_1|^2)^2 + 2 |a_0 a_{-1}^* - a_0^* a_1|^2 \sep
	\sigma = |a_1|^2 - |a_{-1}|^2 ~.
\end{align}
In terms of the action, the superposition of extremally polarized vector solitons is possible because of a $\rm{U(1)}$ symmetry for each component of the field in orthonormal polarization bases \cite{Jain:2021pnk}. Since the energy expression \eqref{soliton_energy} does not explicitly depend on the spin density in this case, solitons with different polarization states have the same energy at a given particle number and thus are equally favored energetically. 

In the presence of a polarization-dependent SI coupling $\til\lambda_\sigma$, it would not be possible to superpose the base polarization vectors as in \eqref{polarization_superposition} to create a partially polarized soliton. To understand how the energy of solitons varies with different polarization states, consider a small variation in the SI coupling $\delta_\lambda\equiv \delta\til\lambda_\sigma/\til\lambda_\sigma$. The resulting change in the soliton profile, while preserving the total particle number, will be on the order of $|\delta\til f| \sim \epsilon |\delta_\lambda|$, where $\epsilon$ represents the small dimensionless quantities \eqref{small_parameter1} and \eqref{small_parameter2} that are used to derive the nonrelativistic effective field theory. Under the leading-order approximation, the energy difference between solitons with different polarization, according to \eqref{soliton_energy}, is given by $\delta \til E_s = \int d^3\til x ~ \frac{1}{16} \til\lambda_\sigma \til f^4 \delta_\lambda$. For attractive SIs with $\til\lambda_\sigma<0$, a nonzero $\sigma$ results in a larger $\til\lambda_\sigma$ and thus higher energy compared to the case where $\sigma=0$; the opposite result occurs for $\til\lambda_\sigma>0$.

\subsection{Impacts of nonminimal gravitational interactions}
\label{sec:impacts_ngi}
Compared to the SIs, the attractive or repulsive nature of the force induced by the NGIs is not obvious due to the gradient dependence in the coupling. To assess the effect, one approach is to examine whether the mass density $\til \rho$ is enhanced or reduced after incorporating the modification due to $\til\xi$. Assuming that soliton profiles can be approximated by a Gaussian function $\til f(r) = \til C e^{-\til r^2/\til R_e^2}$, the Laplacian of $\til\rho$, which becomes $4(4\til r^2-3\til R_e^2) \til f^2/\til R_e^4$, is negative within the soliton core but reverses sign in the outer regions. Therefore, a positive $\til \xi$ would reduce the mass density and induce a repulsive force within the core of solitons, while causing opposite effects in the outer regions.

Apart from the new force, the NGIs imply a critical amplitude of solitons beyond which soliton solutions do not exist, regardless of the sign of $\til\xi$. To see this, we can think of \eqref{soliton_profile_eq_v1} as an equation of motion for a ball rolling down a hill \cite{Zhang:2020bec},
\begin{align}
	\pd_{\til r}^2 \til f + \frac{2}{\til r}\pd_{\til r} \til f = -U_\rm{eff}' (\til f) ~,
\end{align}
where $\til f$ and $\til r$ are regarded as location and time, $(2/\til r)\pd_{\til r} \til f$ is a time-dependent damping term, and $U_\rm{eff}$ is an effective potential. Imagining that somehow we know the form of $U_\rm{eff}(\til f)$, a soliton solution can be thought of as a ball rolling from $\til f = \til f_0$ at time $\til r=0$ and reaching $\til f = 0$ at infinity time, where $\til f_0$ is the central amplitude of the soliton. There is a requirement for the form of the effective potential: Without the damping term, e.g., in 1-dimensional space, the effective potential must satisfy $U_\rm{eff}(\til f_0) = U_\rm{eff}(0)$ and has a local minimum between $\til f_0$ and $0$; the existence of the damping term requires the ``initial energy'' $U_\rm{eff}(\til f_0)$ to be larger than $U_\rm{eff}(0)$. Under these conditions, the initial acceleration,
\begin{align}
	\label{initial_acceleration_ngi}
	\pd_{\til r}^2 \til f \Big|_{\til r=0} = \frac{\[\til\gamma - 8\til\xi \] \til f^3 + 8\til f \til\Psi}{4\[1 - 4\til\xi^2 \til f^2 \]} \Bigg|_{\til r=0} ~,
\end{align}
should be negative. Now, consider that we are trying to find a soliton solution with larger and larger amplitudes. At small amplitudes, the numerator (denominator) of \eqref{initial_acceleration_ngi} is negative (positive). As we increase the amplitude, the denominator vanishes at the critical amplitude
\begin{align}
	\label{critical_amp}
	\til f_{0}^\rm{crit} = \frac{1}{2|\til \xi|} ~,
\end{align}
beyond which the rolling ball scenario fails and no soliton solutions could be found.\footnote{This critical amplitude was not identified in reference \cite{Zhang:2023fhs}, where the assumptions $\til R_s^2\gg \til \xi$ and $\til \rho_\xi\approx \til\rho$ are used to study the mass-radius relation of solitons. However, as we will see shortly, the assumption $\til\rho_\xi\approx\til\rho$ breaks down at some point.}

As demonstrated in previous numerical studies \cite{Chen:2020cef, Mocz:2023adf}, solitons continue to grow after their formation, and at some point, SIs begin to play a significant role. It has been observed that while solitons can become increasingly compact under the influence of repulsive SIs, attractive SIs destabilize solitons when their energy becomes comparable to the gravitational energy \cite{Chen:2020cef}. However, the presence of the NGIs could change the scenario: Solitons are predicted to collapse if their amplitudes reach the critical value \eqref{critical_amp}. In the next section, we will explore how this is reflected in the mass-radius relation of solitons. The growth of solitons in dark matter halos is investigated numerically in a subsequent paper \cite{Chen:2024pyr}.

\section{Soliton's mass-radius relation, stability, and profiles}
\label{sec:mr_relation}
Since stable solitons represent the ground state of a collection of particles with a fixed particle number, their formation and evolution are not sensitive to initial conditions. Insights into their dynamical properties can therefore be gained by examining their stationary properties, such as the relationship between two observables, mass and radius. This relation turns out to have close connection to the stability of solitons.

\begin{figure}
	\centering
	\includegraphics[width=0.5\linewidth]{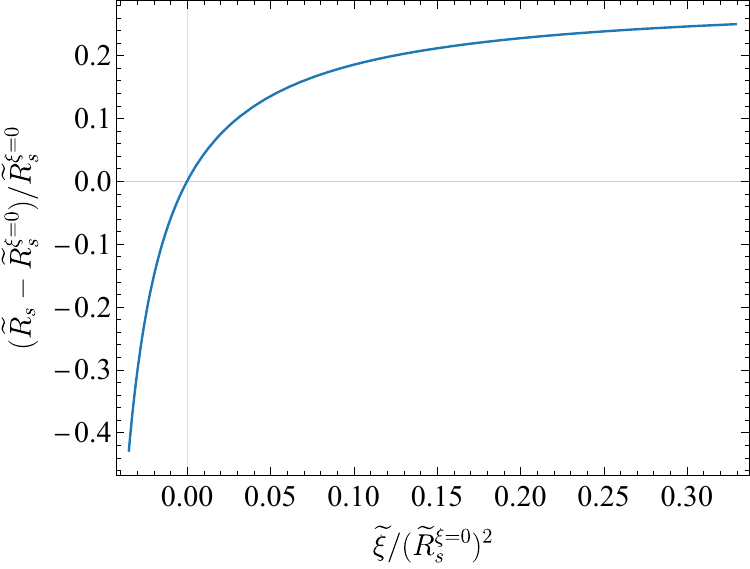}
	\caption{Fractional difference between the mass radius $\til R_s$ and the particle radius $\til R_s^{\xi=0}$, which enclose $95\%$ of the total mass or particle number of solitons.}
	\label{fig:radiuscorrection}
\end{figure}

To find an analytical expression for the mass-radius relation, we can express the soliton energy \eqref{soliton_energy} as
\begin{align}
	\label{soliton_energy_mr}
	\til E_s \propto \frac{c_0}{2} \til M_s \( \frac{d\til R_s}{d \til t} \)^2 + c_1 \frac{\til M_s}{\til R_s^2} - \frac{2\til M_s^2}{\til R_s} + \frac{2 c_2}{3} \frac{\til\gamma \til M_s^2}{\til R_s^3} - \frac{2c_3^2 \til\xi^2\til M_s^2}{5 \til R_s^5} ~,
\end{align}
where the first two terms correspond to the gradient energy,\footnote{The kinetic term comes from the derivative of the phase of the field, which is $\sim mv r \sim m\dot{R_s} r$ in the plane-wave approximation. As far as I know, this term was first pointed out in \cite{Chavanis:2011zi, Chavanis:2011zm} by working with fluid equations and assuming Gaussian density profiles for scalar dark matter.} the third term is the gravitational energy, the last two terms represent the energy due to SIs and NGIs, and $c_i$ are positive coefficients that generally depend on the sign and/or the ratio of $\til\gamma$ to $\til\xi$ because of the scaling symmetry \eqref{nondimensionalize}. To write in this way, I have interchanged the use of the particle number radius $\til R_s^{\xi=0}$, defined as enclosing $95\%$ of the total particle number, with the mass radius $\til R_s$. While this substitution is not exact, we can see in figure \ref{fig:radiuscorrection} that their difference is small when $|\til\xi|/\til R_s^2 \lesssim 0.02$.\footnote{To estimate the difference between the mass radius and the particle radius, we note that at large radius the soliton profile satisfies
\begin{align*}
	\til f (\til r) \sim \( \frac{\til\mu \til M_s^2}{2\pi^2} \)^{1/4} \frac{e^{-\sqrt{2\til \mu} \til r}}{\til r} \sep
	\int_{\til R_{s}}^{\infty} \til \rho_\xi ~4\pi \til r^2 d\til r \simeq 0.05 \til M_s ~.
\end{align*}
Linearizing the equation in the fractional difference in radius, we find that $\til R_s$ and $\til R_s^{\xi=0}$ differ by a factor $\lesssim 10\%$ for $|\til\xi|/\til R_s^2 \leq 0.02$, as depicted in figure \ref{fig:radiuscorrection}.}
The equation \eqref{soliton_energy_mr} can be interpreted as a Hamiltonian of a ball with mass $c_0 \til M_s$ rolling in a potential. By minimizing the energy \eqref{soliton_energy_mr} at a fixed particle number (or equivalently, a fixed mass), we find the mass-radius relation of solitons to be
\begin{align}
	\label{soliton_mr_relation}
	\til M_\rm{I} = \frac{c_1 \til R_\rm{I}^3 }{\til R_\rm{I}^4 - c_2 \frac{\til\gamma}{|\til\gamma|+|\til\xi|} \til R_\rm{I}^2 + c_3^2 \( \frac{\til\xi}{|\til\gamma|+|\til\xi|} \)^2} ~,
\end{align}
where the scale-invariant mass and radius are defined as $\til M_\rm{I} \equiv (|\til\gamma| + |\til\xi|)^{1/2} \til M_s$ and $\til R_\rm{I} \equiv (|\til\gamma| + |\til\xi|)^{-1/2} \til R_s$. At large radii, this relation should converge to that for $\til\gamma=\til\xi=0$, making $c_1$ a constant independent of the sign or the ratio of $\til\gamma$ to $\til\xi$. For solitons to be stable and correspond to states with minimized energy, the second derivative of the energy \eqref{soliton_energy_mr} with respect to the soliton radius should be positive,
\begin{align}
	\frac{d^2\til E_s}{d\til R_s^2} \Bigg|_{\til M_s} \propto \frac{2 \til M_s^2}{\til R_s^7} \[ c_2 \til\gamma \til R_s^2 + \frac{F^2}{2\MP^2} ( \til R_s^4 - 3c_3^2\til\xi^2 ) \] \propto -\frac{2 c_1}{\til R_s^3} \frac{d\til M_s}{d\til R_s} > 0 ~.
\end{align}
Therefore, a turnover point in the mass-radius relation indicates a change of stability of solitons. Solitons are stable if $d\til M_s/d\til R_s<0$. For the rest of this section, we will focus on some special cases and numerically analyze the soliton profiles and the mass-radius relation. Numerical calculations are carried out using the Mathematica package ``DMSolitonFinder'', see appendix \ref{sec:mathematica} for details.

\subsection{Minimal scenario: $\til\lambda_\sigma=0$ and $\til\xi=0$}
The simplest case we can consider is one without SIs and NGIs. In this scenario, the mass-radius relation \eqref{soliton_mr_relation} simplifies to
\begin{align}
	\label{soliton_mr_minimal}
	\til M_s \til R_s = c_1 ~,
\end{align}
where $c_1 = 98.5$ is determined numerically. Corresponding to the approximate density profile in \cite{Schive:2014dra}, we can find an excellent approximation for the soliton profile
\begin{align}
	\label{soliton_profile_approx}
	\til f(\til r) = \frac{f_0^2}{(1+0.0529 ~f_0^2 \til r^2)^{4}} ~,
\end{align}
where $f_0$ is the central field amplitude. This profile yields a relative error of less than $0.5\%$ within the soliton radius when compared to the numerical profile with the same central amplitude. Numerically, solitons with a central amplitude $\til f(0) = f_0^2$ have $\til M_s = 25.9 f_0$, $\til R_s=3.80/f_0$ and $\til \mu=0.692 f_0^2$. For $f_0\sim 1$, the nonrelativistic effective field theory breaks down as the small parameters defined in \eqref{small_parameter1} are order unity.\footnote{In term of dimensional quantities, the nonrelativistic effective field theory breaks down for solitons with $M_s\gtrsim 50 \MP^2/m$ or $R_s \lesssim 4/m$, in agreement with \cite{Salehian:2021khb}.} Beyond this point, one must refer to the full relativistic theory to characterize the nonlinear dynamics of solitons.

\subsection{Self-interactions: $\til\lambda_\sigma\neq 0$ and $\til\xi=0$}
\begin{figure}
	\centering
	\includegraphics[width=0.5\linewidth]{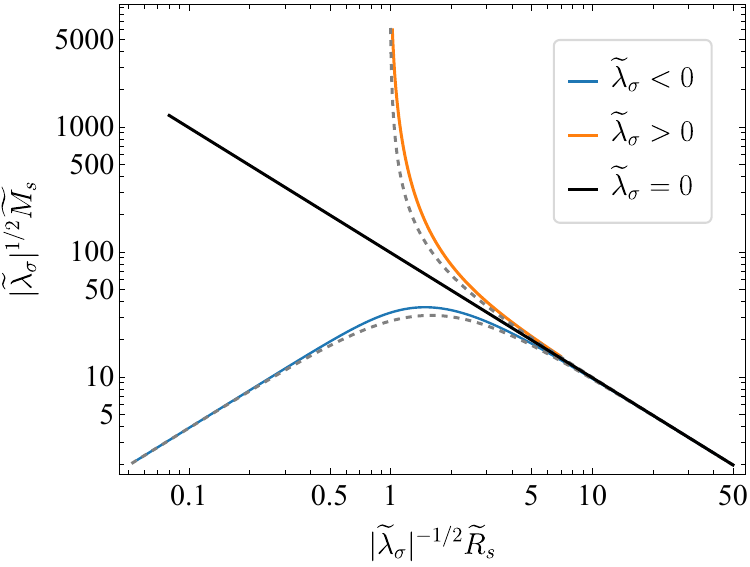}
	\caption{Mass-radius relation for solitons with SIs and negligible NGIs. Solid lines are obtained numerically and corresponding gray dashed lines are analytical approximations based on \eqref{soliton_mr_si}.}
	\label{fig:mrrelationsi}
\end{figure}

If $\lambda$ satisfies \eqref{lowlimit_lambda_xi}, such as QCD axions with $\lambda \simeq -10^{-52} [m_a/(10^{-5} \rm{eV})]^4$, SIs play a nonnegligible role in soliton dynamics. In cases where NGIs are negligible and $\til\gamma=\til\lambda_\sigma$, the mass-radius relation \eqref{soliton_mr_relation} becomes
\begin{align}
	\label{soliton_mr_si}
	|\til\lambda_\sigma|^{1/2} \til M_s = \frac{c_1 ( |\til\lambda_\sigma|^{-1/2} \til R_s )}{( |\til\lambda_\sigma|^{-1/2} \til R_s )^2 - c_2 \rm{sgn}[\til\lambda_\sigma] } ~,
\end{align}
where $\rm{sgn}[\cdots]$ is the sign function and $c_2$ depends only on the sign of $\til\lambda_\sigma$. Note that this relation is also applicable to scalar and complex vector solitons if we replace $\til\lambda_\sigma$ with $\til\lambda$, consistent with the results for scalar solitons \cite{Chavanis:2011zi, Chavanis:2011zm, Chavanis:2022fvh}. The mass-radius relation is shown in figure \ref{fig:mrrelationsi}.

For $\til\lambda_\sigma<0$, the SI term dominates over the gravity term in \eqref{soliton_energy_mr} when the soliton radius is small. Numerical calculations yield $c_2=2.53$. According to \eqref{soliton_mr_si}, the maximum mass of solitons occurs at
\begin{align}
	\label{soliton_marginal_stable}
	|\til\lambda_\sigma|^{-1/2} \til R_s^\rm{min} = c_2^{1/2} \sep
	|\til\lambda_\sigma|^{1/2} \til M_s^\rm{max} = \frac{c_1}{2} \( \frac{1}{c_2} \)^{1/2} ~,
\end{align}
which well approximate the numerical results $|\til\lambda_\sigma|^{-1/2} \til R_s^\rm{min} = 1.49$ and $|\til\lambda_\sigma|^{1/2} \til M_s^\rm{max} = 36.0$. At this point, $d\til M_s/d\til R_s=0$, indicating the smallest radius of stable solitons. For a dark matter clump with $\til M > \til M_s^\rm{max}$, it would collapse to a black hole unless some other processes such as virialization halt the shrinking. To put it another way, the collapse occurs when the dark matter density reaches the core density of the soliton with the maximum mass
\begin{align}
	|\til\lambda_\sigma|^2 \til \rho_\rm{max} = 33.2 ~,
\end{align}
in agreement with \cite{Chen:2020cef}. A dark matter clump initially in the unstable branch of the mass-radius relation may either collapse into a black hole or relax towards the stable branch.

For $\til\lambda_\sigma>0$, there exists a minimum soliton radius
\begin{align}
	\til R_s^\rm{min} = \sqrt{c_2 \til\lambda_\sigma} ~.
\end{align}
To determine the value of $c_2$, we can apply the Thomas-Fermi approximation to solve the soliton profile. In this approximation, the quantum pressure (i.e., the gradient term) is neglected, and the profile equation \eqref{soliton_profile_eq_v2} becomes
\begin{align}
	\label{soliton_eqf_thomas_fermi}
	\til\nabla^2 \til\rho + \frac{8}{\til\lambda_\sigma} \til\rho = 0 ~.
\end{align}
The solution to this equation is
\begin{align}
	\til \rho(\til r) = \frac{\pi M_s}{4\pi \til R^2 \til r} \sin\( \frac{\pi \til r}{\til R} \) \sep
	\til R = \frac{\pi\sqrt{2\til\lambda_\sigma}}{4} ~,
\end{align}
in agreement with the result for scalar solitons \cite{Chavanis:2011zi, Chavanis:2011zm, Chavanis:2022fvh}. This implies that solitons with a radius $\til R_s^\rm{min}$ are localized in space with a size $\til R$. The radius enclosing $95\%$ of the total mass is related to $\til R$ through $\til R_s^\rm{min}=0.895 \til R$, leading to the determination of $c_2 = 0.988$.

\subsection{Nonminimal gravitational interactions: $\til\lambda_\sigma=0$ and $\til\xi\neq 0$}
\begin{figure}
	\centering
	\includegraphics[width=0.5\linewidth]{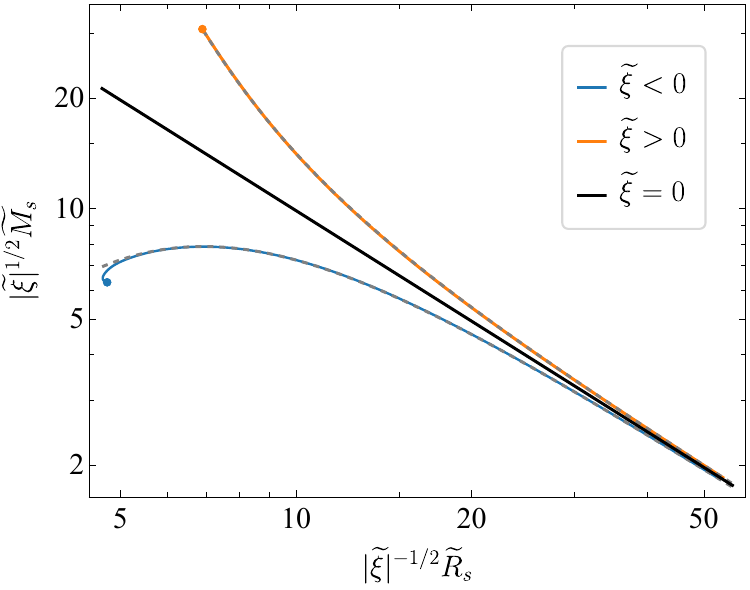}
	\caption{Mass-radius relation for solitons with NGIs. Solid lines are obtained numerically and corresponding gray dashed lines are predictions by \eqref{soliton_mr_ngi}. Two dots represent solitons with the critical amplitude \eqref{critical_amp}, beyond which no soliton solutions exist. For $\til\xi<0$, the maximum mass and smallest radius of stable solitons correspond to the turnover point described by \eqref{soliton_turnover_ngi}. For $\xi>0$, the maximum mass and smallest radius of solitons are given by \eqref{critical_point_ngi}.}
	\label{fig:mrrelationngi}
\end{figure}

If $\xi$ satisfies \eqref{lowlimit_lambda_xi}, NGIs may play a significant role in soliton dynamics. Neglecting SIs, the mass-radius relation \eqref{soliton_mr_relation} becomes
\begin{align}
	\label{soliton_mr_ngi}
	|\til \xi|^{1/2}\til M_s = \frac{c_1 (|\til \xi|^{-1/2} \til R_s)^3 }{ (|\til \xi|^{-1/2} \til R_s)^4 - 16 c_2 \rm{sgn}[\til\xi] (|\til \xi|^{-1/2} \til R_s)^2 + c_3^2 } ~,
\end{align}
where $c_2$ depends only on the sign of $\til\xi$. As noted in section \ref{sec:impacts_ngi}, there exists a critical soliton solution whose amplitude is given by \eqref{critical_amp}. To estimate when this becomes relevant, we may use the approximate soliton profile \eqref{soliton_profile_approx} to find the critical soliton mass and radius,
\begin{align}
	\label{critical_mr}
	|\til\xi|^{1/2} \til M_s^\rm{crit} \sim 18.3 \sep
	|\til\xi|^{-1/2} \til R_s^\rm{crit} \sim 
	\begin{cases}
		3.41 \sep \text{for } \til\xi<0 \\
		6.23 \sep \text{for } \til\xi>0
	\end{cases} ~.
\end{align}
The mass-radius relation is shown in figure \ref{fig:mrrelationngi}.

\begin{figure}
	\centering
	\begin{minipage}{0.47\linewidth}
		\includegraphics[width=\linewidth]{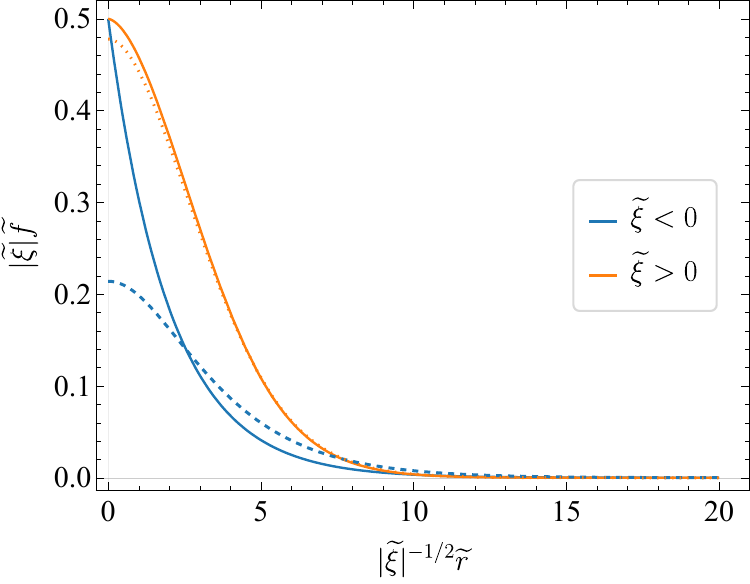}
	\end{minipage}
	\quad
	\begin{minipage}{0.495\linewidth}
		\includegraphics[width=\linewidth]{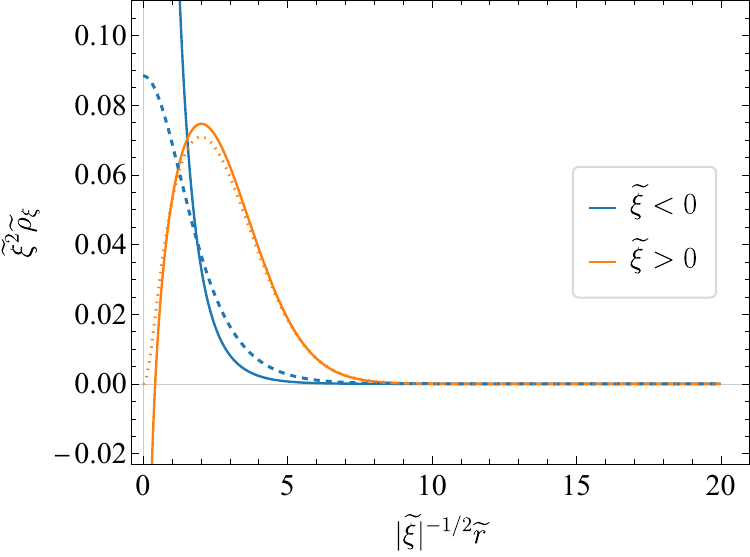}
	\end{minipage}
	\caption{Field (left) and density (right) profiles of solitons. The solid lines represent the critical solitons, characterized by central amplitudes approaching \eqref{critical_amp} and central densities diverging to $\pm \infty$. The (blue) dashed lines correspond to the marginally stable soliton for $\til\xi<0$, with the mass given by \eqref{soliton_turnover_ngi}. The (orange) dotted lines represent the soliton with vanished density $\rho_\xi$ at the center.}
	\label{fig:ngicriticalprofile}
\end{figure}

For $\til\xi<0$, the maximum mass of solitons occurs at
\begin{align}
	\label{soliton_turnover_ngi}
	|\til\xi|^{-1/2} \til R_s^\rm{min} = \[ 8 c_2 + (64 c_2^2 + 3c_3^2)^{1/2} \]^{1/2} \sep
	|\til\xi|^{1/2} \til M_s^\rm{max} = \frac{c_1 [ 8 c_2 + (64c_2^2 + 3 c_3^2)^{1/2} ]^{3/2} }{4 c_3^2 + 32 c_2 [8 c_2 + (64c_2^2 + 3 c_3^2)^{1/2}] } ~.
\end{align}
Due to the critical amplitude \eqref{critical_amp}, it is not possible to take the small-radius limit to determine the numerical values of $c_2$ or $c_3$, as we did for the case with SIs. Instead, to make the analytical approximation \eqref{soliton_turnover_ngi} agree with the numerical results $|\til\xi|^{-1/2} \til R_s = 6.94$ and $|\til\xi|^{1/2} \til M_s = 7.86$, we can require $c_2=2.13$ and $c_3=15.0$. Note that the analytical approximation does not capture the behavior near the critical point, indicated as the solid blue point in figure \ref{fig:mrrelationngi}, where the amplitude of solitons approaches \eqref{critical_amp} and the soliton profile becomes sharply peaked at the center (see figure \ref{fig:ngicriticalprofile}). Analogous to the case with SIs, a dark matter clump with $\til M>\til M_s^\rm{max}$ would collapse to a black hole unless some other processes stop the contraction. The collapse happens when the dark matter density reaches the core density of the soliton with the maximum mass
\begin{align}
	|\til\xi|^{2} \til \rho_\xi^\rm{max} = 0.0884 ~.
\end{align}
In this particular case, the critical soliton resides in the unstable branch of the mass-radius relation and is thus not observable. More generally, in the presence of both SIs and NGIs with $\til\gamma<0$ and $\til\xi<0$, one should compare the marginally stable point \eqref{soliton_marginal_stable} due to $\til\gamma$ and the critical point \eqref{critical_mr} due to $\til\xi$ in order to estimate the maximum mass and the minimum radius of solitons.

For $\til\xi>0$, the mass and radius of the critical soliton are given by
\begin{align}
	\label{critical_point_ngi}
	\til\xi^{1/2} \til M_s^\rm{max} = 30.8 \sep
	\til\xi^{-1/2} \til R_s^\rm{min} =  6.91 ~.
\end{align}
To make the prediction of \eqref{soliton_mr_ngi} agree with \eqref{critical_point_ngi}, we can choose $c_2=1.60+0.00131 c_3^2$. By varying the values of $c_3$, one can show that $c_3=20.3$ and thus $c_2=2.14$ provide an excellent approximation for the numerical mass-radius relation, as shown in figure \ref{fig:mrrelationngi}. When the dark matter density reaches the positive peak of the density profile for solitons corresponding to \eqref{critical_point_ngi}, 
\begin{align}
	|\til\xi|^{2} \til \rho_\xi^\rm{max} = 0.0746 ~,
\end{align}
the region will collapse and relativistic effects will become important.

Figure \ref{fig:ngicriticalprofile} displays the field and density profiles of solitons for $\til\xi<0$ (blue) and $\til\xi>0$ (orange). The (blue) dashed lines are the profiles of marginally stable soliton for $\til\xi<0$. The solid lines correspond to the critical solitons with a central amplitude approaching \eqref{critical_amp}. As we see in the plot, while the density $\til\rho_\xi$ remains positive for $\til\xi<0$, it can become negative in the innermost core of solitons for $\til\xi>0$. The central density for the critical solitons approaches $\pm \infty$ since the second derivative of the profile diverges. Additionally, the (orange) dotted lines depict the soliton with vanished central density, indicating that ``repulsive'' gravity with negative $\til\rho_\xi$ can only occur in the innermost region of compact solitons, whose amplitudes are close to the critical amplitude.

\subsection{Gradient-dependent self-interactions: $\til\gamma=0$ and $\til\xi\neq 0$}
\begin{figure}
	\centering
	\includegraphics[width=0.5\linewidth]{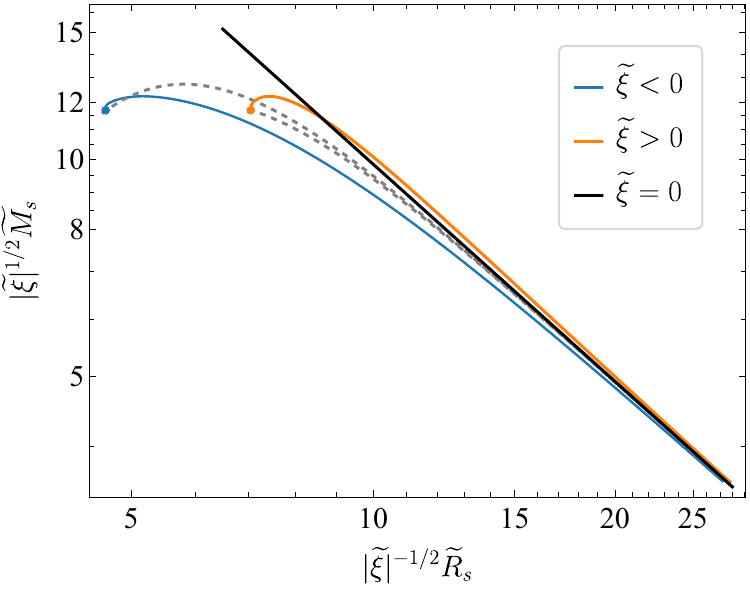}
	\caption{Mass-radius relation for solitons with gradient-dependent SIs. The notation follows that of figure \ref{fig:mrrelationngi}.}
	\label{fig:mrrelationgradientsi}
\end{figure}

In the presence of both SIs and NGIs, it is useful isolate the impact of the gradient-dependent part by considering the case where $\til\gamma=0$. This simplifies the mass-radius relation to
\begin{align}
	\label{soliton_mr_gradient_si}
	|\til\xi|^{1/2}\til M_s = \frac{c_1 (|\til\xi|^{-1/2} \til R_s)^3 }{ (|\til\xi|^{-1/2} \til R_s)^4 + c_3^2 } ~,
\end{align}
where $c_3$ depends on the sign of $\til\xi$. Solving for soliton profiles with central amplitudes approaching the critical value \eqref{critical_amp}, we find the critical mass and radius of solitons,
\begin{align}
	\label{soliton_critical_gradient_si}
	|\til\xi|^{1/2} \til M_s = 11.7 \sep
	|\til\xi|^{-1/2} \til R_s = 
	\begin{cases}
		4.65 \sep \text{for } \til\xi<0 \\
		7.04 \sep \text{for } \til\xi>0
	\end{cases} ~.
\end{align}
Interestingly, the critical mass is identical for $\til\xi<0$ and $\til\xi>0$, as it is determined by the soliton profile, which is independent of the sign of $\til\xi$. To make the prediction of \eqref{soliton_mr_gradient_si} align with \eqref{soliton_critical_gradient_si}, we set $c_3=19.5$ for $\til\xi<0$ and $c_3=21.9$ for $\til\xi>0$. The mass-radius relation is shown in figure \ref{fig:mrrelationgradientsi}.

\section{Conclusions}
In this work, we aim to provide a unified perspective on solitons in ultralight scalar and vector dark matter, which are known by various names such as boson/Bose/soliton stars, axion stars, oscillatons, Q-balls, oscillons/I-balls/quasi-breathers, Proca stars, Proca Q-balls, and vector oscillons depending on specific contexts. To achieve this, we begin with Lorentz-invariant actions, including the leading SIs and NGIs of dark matter, and derive the nonrelativistic effective field theory following the strategy in \cite{Salehian:2020bon, Salehian:2021khb}. Some of our key findings in the nonrelativistic regime include:
\begin{enumerate}
	\item Real vector dark matter represents the most general scenario, in the sense that the field equations can reduce to the other three cases---where dark matter consists of real scalars, complex scalars, or complex vectors---through a few simple manipulations of variables.
	\item Under the assumption that dark matter fields are even under the $\mathbb{Z}_2$ reflection symmetry, it is challenging to distinguish between real and complex scalar dark matter, and consequently, their solitons, without accounting for relativistic effects or interactions beyond SIs and NGIs.
	\item The quartic SI of real vector dark matter incorporates (macroscopic) spin-spin interactions, which break the $\rm{U}(1)$ symmetry in each component of the field in orthonormal polarization bases and prohibits the superposition of extremally polarized solitons \cite{Jain:2021pnk}. Moreover, the NGIs of dark matter introduce gradient-dependent SIs and modify Poisson's equation for the gravitational potential.
\end{enumerate}
Based on these findings, we demonstrate that dark matter solitons in these four scenarios share universal nonrelativistic properties, such as conserved charges, mass-radius relations, stability, and profiles. Specifically, we study the mass-radius relation of real vector solitons for several benchmark examples, including the case with NGIs and purely gradient-dependent SIs (a combination of the SIs and NGIs). The stability of solitons against perturbations is ensured if $dM_s/dR_s<0$. Here, numerical calculations are performed using the Mathematica package ``DMSolitonFinder'', which dynamically adjusts the boundary conditions and the spatial range of solutions until localized solutions are found with the requested precision and accuracy. See appendix \ref{sec:mathematica} for more details.

Dark matter solitons could form through gravitational Bose-Einstein condensation in the kinetic regime \cite{Levkov:2018kau, Chen:2023bqy, Jain:2023ojg, Jain:2023tsr}. After their formation, solitons continue to grow and at some stages SIs start to become important \cite{Chen:2020cef}. For example, solitons with attractive SIs (e.g., axion stars) collapse upon reaching a critical mass when their SI energy becomes comparable to their gravitational energy; while those with repulsive SIs keep growing without collapsing. The presence of the NGIs could change the story: Solitons are predicted to collapse if their amplitudes reach the critical value \eqref{critical_amp}, regardless of the sign of the coupling. This imposes an upper bound on solitons' mass at the level of $\sim 40 \MP^2/[ m|\xi|^{1/2} ]$, thereby affecting their mass distribution and phenomenology. Moreover, the NGIs modify Poisson's equation for the gravitational potential. With a NGI coupling $\xi<0$, solitons become more compact compared to those without the NGIS. On the other hand, if $\xi>0$, a core with ``repulsive gravity'' could form for solitons whose amplitudes are close to the critical value \eqref{critical_amp}. This may lead to novel signals in gravitational lensing, which will be explored in future work.

In summary, dark matter solitons constituting real or complex, scalar or vector particles can be regarded as related objects. We explore their mass-radius relation and stability, and discuss the impacts of SIs and NGIs. The novel properties of solitons due to NGIs could have interesting phenomenological implications that are worth investigating in future studies.

\acknowledgments

I would like to thank Mustafa A. Amin, Andrew J. Long, Enrico D. Schiappacasse, and Luca Visinelli for helpful discussions and suggestions on the manuscript. This work is supported in part by the National Natural Science Foundation of China (NSFC) through grant No. 12350610240.

\appendix

\section{Conserved charges}
\label{sec:conserved_charges}
In this appendix, symmetries preserved by the nonrelativistic actions \eqref{nr_action_gravity} and \eqref{nr_action_re_vec} and the associated conserved charges are identified. For simplicity, let us first consider a nonexpanding universe, i.e., setting $a=1$, $H=0$, and $\bar\rho_\xi=0$. At the end I will discuss the impact of the expansion of the universe.

In a nonexpanding universe, the nonrelativistic action is invariant under an infinitesimal spacetime translation $x^\mu \rightarrow x^\mu + a^\mu$. The associated Noether's current is the energy-momentum tensor, and here we are interested in the $0\nu$ component, which is given by
\begin{align}
	\t T{^0_\nu} = -i\psi_j^* \pd_\nu \psi_j + \delta^0_\nu \cal L .
\end{align}
The energy density of the system is $-\t T{^0_0}$ and thus the total energy is
\begin{align}
	\label{def_energy}
	E = \int d^3x \[ \frac{1}{2m}\pd_j\psi_i\pd_j\psi_i^* + \frac{1}{2}\Phi\rho_\xi + \frac{\lambda}{48m^2} \( |\b\psi\cdot\b\psi|^2 + 2|\b\psi|^4 \) \] ~,
\end{align}
where the gravitational energy is
\begin{align}
	\label{def_gravitational_energy}
	\int d^3x ~\frac{1}{2}\Phi\rho_\xi = - \frac{G}{2} \int d^3x d^3y ~\frac{\rho_\xi(\b x) \rho_\xi(\b y)}{|\by-\bx|} ~.
\end{align}
To retain the Newton's law for gravity, we may define the mass of a field configuration as
\begin{align}
	\label{def_mass}
	M = \int d^3x ~\rho_\xi ~.
\end{align}

The nonrelativistic action is invariant under an infinitesimal global $\rm{U}(1)$ transformation $\psi_i \rightarrow \psi_i - i \alpha \psi_i$ and $\psi_i^* \rightarrow \psi_i^* + i \alpha \psi_i^*$. The associated conserved charge may be identified as the particle number
\begin{align}
	\label{def_particle_number}
	N = \int d^3x ~ n ~,
\end{align}
where $n\equiv \rho/m = |\b\psi|^2$ is the particle number density.

The nonrelativistic action is invariant under an infinitesimal global $\rm{SO}(3)$ transformation $\psi_j\rightarrow \psi_j - \omega_{jk}\psi_k$, where $\omega_{jk} = -\omega_{kj} \equiv \ve_{ijk}\theta_i$ and $\theta_i$ is the rotation angle. The associated conserved charge can be identified as the internal spin
\begin{align}
	\label{def_spin}
	S_i = \int d^3x ~ \cal S_i ~,
\end{align}
where $\cal S_i \equiv i\ve_{ijk} \psi_j \psi_k^*$ is the spin density. It thus turns out that the $\lambda$-dependent term in \eqref{def_energy} depends on spin,
\begin{align}
	\label{spin_interaction}
	\frac{\lambda}{48m^2}\( |\b\psi\cdot\b\psi|^2 + 2|\b \psi|^4 \) = \frac{\lambda}{48m^2} \( 3|\b\psi|^4 - \b{\cal S}^2 \) ~,
\end{align}
where $|\b{\cal S}|^2 = |\b\psi|^4 - |\b\psi\cdot\b\psi|^2$. As discussed in the main text, the existence of the spin-spin interaction tends to increase (decrease) the total energy for attractively (repulsively) self-interacting real vector dark matter.

The nonrelativistic action is also invariant under an infinitesimal space rotation $x_j \rightarrow x_j - \omega_{jk} x_k$, for which case the field $\psi_j$ changes as $\psi_j (x) \rightarrow \[\delta_j^k - \frac{i}{2} \omega_{mn} \t{(J^{mn})}{_j^k} \] \psi_k(x+\omega x)$, where $\t{(J^{mn})}{_j^k}=-i(\delta^m_j \eta^{nk} - \eta^{mk}\delta^n_j)$ and $\eta^{\mu\nu}$ is the Minkowski metric. The associated conserved charge can be identified as the total angular momentum
\begin{align}
	J_i = L_i + S_i = \int d^3x ~\cal L_i + S_i~,
\end{align}
where $\cal L_i \equiv \ve_{ijk} x^j T^{0k} = -i\ve_{ijk} x^j \psi_l^* \pd_k \psi_l$ is the orbital angular momentum density. The total spin and orbital angular momentum are conserved separately.

In an expanding universe, the time translation symmetry is broken and the energy \eqref{def_energy} is no longer conserved. However, the $\rm{U}(1)$ and $\rm{SO}(3)$ symmetry in field configurations and space are still preserved. The foregoing definitions of mass, particle number, and angular momentum now should be regarded as comoving and appropriate factors in terms of the scale factor $a(t)$ should be taken into account for quantities in physical coordinates.

\section{Mathematica package: DMSolitonFinder}
\label{sec:mathematica}
In this work, I develop the Mathematica package ``DMSolitonFinder'', available on \url{https://github.com/hongyi18/DMSolitonFinder/}, to solve dark matter soliton profiles in an automated way. To solve soliton profiles, it will dynamically change the boundary conditions and the spatial range of solutions until localized solutions are found under the requested precision and accuracy. For example, soliton solutions with central amplitude $\til f(\til r=0) = 0.01$, particle spin $\sigma=1$, SI coupling $\til\lambda=30$, and NGI coupling $\til\xi=-5$ could be found by using the function \verb|ShootFields|, e.g.,
\begin{verbatim}
     ShootFields[0.01, SI->30, NGI->-5, Spin->1]
\end{verbatim}
The output of \verb|ShootFields| is a list of soliton solutions in the form of $\{ \{\til f(\til r), \til\Psi(\til r)\}, \til r_f \}$, where $\til r_f$ is the outer boundary of the solutions. The basic algorithm used in \verb|ShootFields| (as appeared in ``DMSolitonFinder'' version 1.0) is described as follows:
\begin{enumerate}
	\item Given the central amplitude $\til f(\til r=0)$ and other optional conditions (e.g., the values of $\til\lambda, \til\xi, \sigma$), solve the soliton profile equation \eqref{soliton_profile_eq_v1} with an initial guess of $\til \Psi(\til r=0)$, which is demanded to be less than the central value of the true solution such that the first local minimum of $\til f(\til r)$ from small to large radii is negative.
	\item Solve the equation \eqref{soliton_profile_eq_v1} with a central value of $\til \Psi$ increased by $d\til\Psi$, where $d\til\Psi$ is a small positive number, until the first minimum of the new $\til f(\til r)$ becomes positive.
	\item Stop the calculation and return the solutions if the values of the new $\til f(\til r)$ near the outer boundary $r_f$ is small enough compared to the central amplitude $\til f(\til r=0)$, which is determined by the option \verb|AmpTolerance| and the default value is $10^{-4}$.
	\item If the first minimum of the new $\til f(\til r)$ is located near the outer boundary $r_f$, then increase the boundary $r_f$ for better convergence of soliton solutions, otherwise revert back to the last $\til \Psi(\til r=0)$ and reduce the value of $d\til\Psi$.
	\item Repeat the step 2--4 until solutions that satisfy \verb|AmpTolerance| are found. Warning messages will be generated if the target solutions could not be found or other problems arise; check the package website for possible solutions.
\end{enumerate}
Once soliton solutions are found, one can calculate the mass, radius, particle number, frequency (chemical potential), and energy of the soliton by using the functions \verb|CalMass|, \verb|CalRadius|, \verb|CalParticleNumber|, \verb|CalFrequency|, and \verb|CalEnergy|. For more examples and details, please visit the package website.

\bibliographystyle{jhep}
\bibliography{ref}

\providecommand{\href}[2]{#2}\begingroup\raggedright\begin{thebibliography}{10}

\bibitem{Hu:2000ke}
W.~Hu, R.~Barkana and A.~Gruzinov, \emph{{Cold and fuzzy dark matter}},
  \href{https://doi.org/10.1103/PhysRevLett.85.1158}{\emph{Phys. Rev. Lett.}
  {\bfseries 85} (2000) 1158}
  [\href{https://arxiv.org/abs/astro-ph/0003365}{{\ttfamily
  astro-ph/0003365}}].

\bibitem{Ferreira:2020fam}
E.~G.~M. Ferreira, \emph{{Ultra-light dark matter}},
  \href{https://doi.org/10.1007/s00159-021-00135-6}{\emph{Astron. Astrophys.
  Rev.} {\bfseries 29} (2021) 7}
  [\href{https://arxiv.org/abs/2005.03254}{{\ttfamily 2005.03254}}].

\bibitem{Hui:2021tkt}
L.~Hui, \emph{{Wave Dark Matter}},
  \href{https://doi.org/10.1146/annurev-astro-120920-010024}{\emph{Ann. Rev.
  Astron. Astrophys.} {\bfseries 59} (2021) 247}
  [\href{https://arxiv.org/abs/2101.11735}{{\ttfamily 2101.11735}}].

\bibitem{Zhang:2023ktk}
H.-Y. Zhang, \emph{{Probing ultralight dark fields in cosmological and
  astrophysical systems}}, Ph.D. thesis, Rice U., 2023.
\newblock \href{https://arxiv.org/abs/2401.00043}{{\ttfamily 2401.00043}}.

\bibitem{Matos:2023usa}
T.~Matos, L.~A. Ure\~na L\'opez and J.-W. Lee, \emph{{Short review of the main
  achievements of the scalar field, fuzzy, ultralight, wave, BEC dark matter
  model}}, \href{https://doi.org/10.3389/fspas.2024.1347518}{\emph{Front.
  Astron. Space Sci.} {\bfseries 11} (2024) 1347518}
  [\href{https://arxiv.org/abs/2312.00254}{{\ttfamily 2312.00254}}].

\bibitem{Navarro:1995iw}
J.~F. Navarro, C.~S. Frenk and S.~D.~M. White, \emph{{The Structure of cold
  dark matter halos}}, \href{https://doi.org/10.1086/177173}{\emph{Astrophys.
  J.} {\bfseries 462} (1996) 563}
  [\href{https://arxiv.org/abs/astro-ph/9508025}{{\ttfamily
  astro-ph/9508025}}].

\bibitem{Schive:2014dra}
H.-Y. Schive, T.~Chiueh and T.~Broadhurst, \emph{{Cosmic Structure as the
  Quantum Interference of a Coherent Dark Wave}},
  \href{https://doi.org/10.1038/nphys2996}{\emph{Nature Phys.} {\bfseries 10}
  (2014) 496} [\href{https://arxiv.org/abs/1406.6586}{{\ttfamily 1406.6586}}].

\bibitem{Levkov:2018kau}
D.~Levkov, A.~Panin and I.~Tkachev, \emph{{Gravitational Bose-Einstein
  condensation in the kinetic regime}},
  \href{https://doi.org/10.1103/PhysRevLett.121.151301}{\emph{Phys. Rev. Lett.}
  {\bfseries 121} (2018) 151301}
  [\href{https://arxiv.org/abs/1804.05857}{{\ttfamily 1804.05857}}].

\bibitem{Veltmaat:2018dfz}
J.~Veltmaat, J.~C. Niemeyer and B.~Schwabe, \emph{{Formation and structure of
  ultralight bosonic dark matter halos}},
  \href{https://doi.org/10.1103/PhysRevD.98.043509}{\emph{Phys. Rev. D}
  {\bfseries 98} (2018) 043509}
  [\href{https://arxiv.org/abs/1804.09647}{{\ttfamily 1804.09647}}].

\bibitem{Mocz:2019pyf}
P.~Mocz et~al., \emph{{First star-forming structures in fuzzy cosmic
  filaments}},
  \href{https://doi.org/10.1103/PhysRevLett.123.141301}{\emph{Phys. Rev. Lett.}
  {\bfseries 123} (2019) 141301}
  [\href{https://arxiv.org/abs/1910.01653}{{\ttfamily 1910.01653}}].

\bibitem{May:2021wwp}
S.~May and V.~Springel, \emph{{Structure formation in large-volume cosmological
  simulations of fuzzy dark matter: impact of the non-linear dynamics}},
  \href{https://doi.org/10.1093/mnras/stab1764}{\emph{Mon. Not. Roy. Astron.
  Soc.} {\bfseries 506} (2021) 2603}
  [\href{https://arxiv.org/abs/2101.01828}{{\ttfamily 2101.01828}}].

\bibitem{Gorghetto:2022sue}
M.~Gorghetto, E.~Hardy, J.~March-Russell, N.~Song and S.~M. West, \emph{{Dark
  photon stars: formation and role as dark matter substructure}},
  \href{https://doi.org/10.1088/1475-7516/2022/08/018}{\emph{JCAP} {\bfseries
  08} (2022) 018} [\href{https://arxiv.org/abs/2203.10100}{{\ttfamily
  2203.10100}}].

\bibitem{Chen:2020cef}
J.~Chen, X.~Du, E.~W. Lentz, D.~J.~E. Marsh and J.~C. Niemeyer, \emph{{New
  insights into the formation and growth of boson stars in dark matter halos}},
  \href{https://doi.org/10.1103/PhysRevD.104.083022}{\emph{Phys. Rev. D}
  {\bfseries 104} (2021) 083022}
  [\href{https://arxiv.org/abs/2011.01333}{{\ttfamily 2011.01333}}].

\bibitem{Amin:2022pzv}
M.~A. Amin, M.~Jain, R.~Karur and P.~Mocz, \emph{{Small-scale structure in
  vector dark matter}},
  \href{https://doi.org/10.1088/1475-7516/2022/08/014}{\emph{JCAP} {\bfseries
  08} (2022) 014} [\href{https://arxiv.org/abs/2203.11935}{{\ttfamily
  2203.11935}}].

\bibitem{Chen:2023bqy}
J.~Chen, X.~Du, M.~Zhou, A.~Benson and D.~J.~E. Marsh, \emph{{Gravitational
  Bose-Einstein condensation of vector or hidden photon dark matter}},
  \href{https://doi.org/10.1103/PhysRevD.108.083021}{\emph{Phys. Rev. D}
  {\bfseries 108} (2023) 083021}
  [\href{https://arxiv.org/abs/2304.01965}{{\ttfamily 2304.01965}}].

\bibitem{Gorghetto:2024vnp}
M.~Gorghetto, E.~Hardy and G.~Villadoro, \emph{{More Axion Stars from
  Strings}},  \href{https://arxiv.org/abs/2405.19389}{{\ttfamily 2405.19389}}.

\bibitem{Lee:1991ax}
T.~Lee and Y.~Pang, \emph{{Nontopological solitons}},
  \href{https://doi.org/10.1016/0370-1573(92)90064-7}{\emph{Phys. Rept.}
  {\bfseries 221} (1992) 251}.

\bibitem{Breit:1983nr}
J.~D. Breit, S.~Gupta and A.~Zaks, \emph{{COLD BOSE STARS}},
  \href{https://doi.org/10.1016/0370-2693(84)90764-0}{\emph{Phys. Lett. B}
  {\bfseries 140} (1984) 329}.

\bibitem{Seidel:1991zh}
E.~Seidel and W.~Suen, \emph{{Oscillating soliton stars}},
  \href{https://doi.org/10.1103/PhysRevLett.66.1659}{\emph{Phys. Rev. Lett.}
  {\bfseries 66} (1991) 1659}.

\bibitem{Kaup:1968}
D.~J. Kaup, \emph{Klein-gordon geon},
  \href{https://doi.org/10.1103/PhysRev.172.1331}{\emph{Phys. Rev.} {\bfseries
  172} (1968) 1331}.

\bibitem{ruffini1969systems}
R.~Ruffini and S.~Bonazzola, \emph{Systems of self-gravitating particles in
  general relativity and the concept of an equation of state}, {\emph{Physical
  Review} {\bfseries 187} (1969) 1767}.

\bibitem{Braaten:2015eeu}
E.~Braaten, A.~Mohapatra and H.~Zhang, \emph{{Dense Axion Stars}},
  \href{https://doi.org/10.1103/PhysRevLett.117.121801}{\emph{Phys. Rev. Lett.}
  {\bfseries 117} (2016) 121801}
  [\href{https://arxiv.org/abs/1512.00108}{{\ttfamily 1512.00108}}].

\bibitem{Schiappacasse:2017ham}
E.~D. Schiappacasse and M.~P. Hertzberg, \emph{{Analysis of Dark Matter Axion
  Clumps with Spherical Symmetry}},
  \href{https://doi.org/10.1088/1475-7516/2018/01/037}{\emph{JCAP} {\bfseries
  01} (2018) 037} [\href{https://arxiv.org/abs/1710.04729}{{\ttfamily
  1710.04729}}].

\bibitem{Visinelli:2017ooc}
L.~Visinelli, S.~Baum, J.~Redondo, K.~Freese and F.~Wilczek, \emph{{Dilute and
  dense axion stars}},
  \href{https://doi.org/10.1016/j.physletb.2017.12.010}{\emph{Phys. Lett. B}
  {\bfseries 777} (2018) 64}
  [\href{https://arxiv.org/abs/1710.08910}{{\ttfamily 1710.08910}}].

\bibitem{Hertzberg:2020dbk}
M.~P. Hertzberg, Y.~Li and E.~D. Schiappacasse, \emph{{Merger of Dark Matter
  Axion Clumps and Resonant Photon Emission}},
  \href{https://doi.org/10.1088/1475-7516/2020/07/067}{\emph{JCAP} {\bfseries
  07} (2020) 067} [\href{https://arxiv.org/abs/2005.02405}{{\ttfamily
  2005.02405}}].

\bibitem{Chavanis:2017loo}
P.-H. Chavanis, \emph{{Phase transitions between dilute and dense axion
  stars}}, \href{https://doi.org/10.1103/PhysRevD.98.023009}{\emph{Phys. Rev.
  D} {\bfseries 98} (2018) 023009}
  [\href{https://arxiv.org/abs/1710.06268}{{\ttfamily 1710.06268}}].

\bibitem{Seidel:1993zk}
E.~Seidel and W.-M. Suen, \emph{{Formation of solitonic stars through
  gravitational cooling}},
  \href{https://doi.org/10.1103/PhysRevLett.72.2516}{\emph{Phys. Rev. Lett.}
  {\bfseries 72} (1994) 2516}
  [\href{https://arxiv.org/abs/gr-qc/9309015}{{\ttfamily gr-qc/9309015}}].

\bibitem{Alcubierre:2003sx}
M.~Alcubierre, R.~Becerril, S.~F. Guzman, T.~Matos, D.~Nunez and L.~A.
  Urena-Lopez, \emph{{Numerical studies of Phi**2 oscillatons}},
  \href{https://doi.org/10.1088/0264-9381/20/13/332}{\emph{Class. Quant. Grav.}
  {\bfseries 20} (2003) 2883}
  [\href{https://arxiv.org/abs/gr-qc/0301105}{{\ttfamily gr-qc/0301105}}].

\bibitem{Coleman:1985ki}
S.~R. Coleman, \emph{{Q Balls}},
  \href{https://doi.org/10.1016/0550-3213(85)90286-X,
  10.1016/0550-3213(86)90520-1}{\emph{Nucl. Phys.} {\bfseries B262} (1985)
  263}.

\bibitem{Kusenko:1997ad}
A.~Kusenko, \emph{{Small Q balls}},
  \href{https://doi.org/10.1016/S0370-2693(97)00582-0}{\emph{Phys. Lett. B}
  {\bfseries 404} (1997) 285}
  [\href{https://arxiv.org/abs/hep-th/9704073}{{\ttfamily hep-th/9704073}}].

\bibitem{Copeland:1995fq}
E.~J. Copeland, M.~Gleiser and H.-R. Muller, \emph{{Oscillons: Resonant
  configurations during bubble collapse}},
  \href{https://doi.org/10.1103/PhysRevD.52.1920}{\emph{Phys. Rev. D}
  {\bfseries 52} (1995) 1920}
  [\href{https://arxiv.org/abs/hep-ph/9503217}{{\ttfamily hep-ph/9503217}}].

\bibitem{Kasuya:2002zs}
S.~Kasuya, M.~Kawasaki and F.~Takahashi, \emph{{I-balls}},
  \href{https://doi.org/10.1016/S0370-2693(03)00344-7}{\emph{Phys. Lett.}
  {\bfseries B559} (2003) 99}
  [\href{https://arxiv.org/abs/hep-ph/0209358}{{\ttfamily hep-ph/0209358}}].

\bibitem{Saffin:2006yk}
P.~M. Saffin and A.~Tranberg, \emph{{Oscillons and quasi-breathers in D+1
  dimensions}},
  \href{https://doi.org/10.1088/1126-6708/2007/01/030}{\emph{JHEP} {\bfseries
  01} (2007) 030} [\href{https://arxiv.org/abs/hep-th/0610191}{{\ttfamily
  hep-th/0610191}}].

\bibitem{Fodor:2008es}
G.~Fodor, P.~Forgacs, Z.~Horvath and A.~Lukacs, \emph{{Small amplitude
  quasi-breathers and oscillons}},
  \href{https://doi.org/10.1103/PhysRevD.78.025003}{\emph{Phys. Rev. D}
  {\bfseries 78} (2008) 025003}
  [\href{https://arxiv.org/abs/0802.3525}{{\ttfamily 0802.3525}}].

\bibitem{Fodor:2008du}
G.~Fodor, P.~Forgacs, Z.~Horvath and M.~Mezei, \emph{{Computation of the
  radiation amplitude of oscillons}},
  \href{https://doi.org/10.1103/PhysRevD.79.065002}{\emph{Phys. Rev. D}
  {\bfseries 79} (2009) 065002}
  [\href{https://arxiv.org/abs/0812.1919}{{\ttfamily 0812.1919}}].

\bibitem{Zhang:2020bec}
H.-Y. Zhang, M.~A. Amin, E.~J. Copeland, P.~M. Saffin and K.~D. Lozanov,
  \emph{{Classical Decay Rates of Oscillons}},
  \href{https://doi.org/10.1088/1475-7516/2020/07/055}{\emph{JCAP} {\bfseries
  07} (2020) 055} [\href{https://arxiv.org/abs/2004.01202}{{\ttfamily
  2004.01202}}].

\bibitem{Zhang:2020ntm}
H.-Y. Zhang, \emph{{Gravitational effects on oscillon lifetimes}},
  \href{https://doi.org/10.1088/1475-7516/2021/03/102}{\emph{JCAP} {\bfseries
  03} (2021) 102} [\href{https://arxiv.org/abs/2011.11720}{{\ttfamily
  2011.11720}}].

\bibitem{Brito:2015pxa}
R.~Brito, V.~Cardoso, C.~A.~R. Herdeiro and E.~Radu, \emph{{Proca stars:
  Gravitating Bose\textendash{}Einstein condensates of massive spin 1
  particles}},
  \href{https://doi.org/10.1016/j.physletb.2015.11.051}{\emph{Phys. Lett. B}
  {\bfseries 752} (2016) 291}
  [\href{https://arxiv.org/abs/1508.05395}{{\ttfamily 1508.05395}}].

\bibitem{Loginov:2015rya}
A.~Y. Loginov, \emph{{Nontopological solitons in the model of the
  self-interacting complex vector field}},
  \href{https://doi.org/10.1103/PhysRevD.91.105028}{\emph{Phys. Rev. D}
  {\bfseries 91} (2015) 105028}.

\bibitem{Adshead:2021kvl}
P.~Adshead and K.~D. Lozanov, \emph{{Self-gravitating Vector Dark Matter}},
  \href{https://doi.org/10.1103/PhysRevD.103.103501}{\emph{Phys. Rev. D}
  {\bfseries 103} (2021) 103501}
  [\href{https://arxiv.org/abs/2101.07265}{{\ttfamily 2101.07265}}].

\bibitem{Jain:2021pnk}
M.~Jain and M.~A. Amin, \emph{{Polarized solitons in higher-spin wave dark
  matter}}, \href{https://doi.org/10.1103/PhysRevD.105.056019}{\emph{Phys. Rev.
  D} {\bfseries 105} (2022) 056019}
  [\href{https://arxiv.org/abs/2109.04892}{{\ttfamily 2109.04892}}].

\bibitem{Zhang:2021xxa}
H.-Y. Zhang, M.~Jain and M.~A. Amin, \emph{{Polarized vector oscillons}},
  \href{https://doi.org/10.1103/PhysRevD.105.096037}{\emph{Phys. Rev. D}
  {\bfseries 105} (2022) 096037}
  [\href{https://arxiv.org/abs/2111.08700}{{\ttfamily 2111.08700}}].

\bibitem{Liebling:2012fv}
S.~L. Liebling and C.~Palenzuela, \emph{{Dynamical Boson Stars}},
  \href{https://doi.org/10.12942/lrr-2012-6}{\emph{Living Rev. Rel.} {\bfseries
  15} (2012) 6} [\href{https://arxiv.org/abs/1202.5809}{{\ttfamily
  1202.5809}}].

\bibitem{Visinelli:2021uve}
L.~Visinelli, \emph{{Boson stars and oscillatons: A review}},
  \href{https://doi.org/10.1142/S0218271821300068}{\emph{Int. J. Mod. Phys. D}
  {\bfseries 30} (2021) 2130006}
  [\href{https://arxiv.org/abs/2109.05481}{{\ttfamily 2109.05481}}].

\bibitem{Mukaida:2014oza}
K.~Mukaida and M.~Takimoto, \emph{{Correspondence of I- and Q-balls as
  Non-relativistic Condensates}},
  \href{https://doi.org/10.1088/1475-7516/2014/08/051}{\emph{JCAP} {\bfseries
  08} (2014) 051} [\href{https://arxiv.org/abs/1405.3233}{{\ttfamily
  1405.3233}}].

\bibitem{Mukaida:2016hwd}
K.~Mukaida, M.~Takimoto and M.~Yamada, \emph{{On Longevity of
  I-ball/Oscillon}}, \href{https://doi.org/10.1007/JHEP03(2017)122}{\emph{JHEP}
  {\bfseries 03} (2017) 122}
  [\href{https://arxiv.org/abs/1612.07750}{{\ttfamily 1612.07750}}].

\bibitem{Birrell:1982ix}
N.~D. Birrell and P.~C.~W. Davies, \emph{{Quantum Fields in Curved Space}},
  Cambridge Monographs on Mathematical Physics. Cambridge Univ. Press,
  Cambridge, UK, 2, 1984,
  \href{https://doi.org/10.1017/CBO9780511622632}{10.1017/CBO9780511622632}.

\bibitem{Weinberg:1995mt}
S.~Weinberg, \emph{{The Quantum theory of fields. Vol. 1: Foundations}}.
  Cambridge University Press, 6, 2005.

\bibitem{callan1970new}
C.~G. Callan~Jr, S.~Coleman and R.~Jackiw, \emph{A new improved energy-momentum
  tensor}, {\emph{Annals of physics} {\bfseries 59} (1970) 42}.

\bibitem{freedman1974energy}
D.~Z. Freedman, I.~J. Muzinich and E.~J. Weinberg, \emph{On the energy-momentum
  tensor in gauge field theories}, {\emph{Annals of Physics} {\bfseries 87}
  (1974) 95}.

\bibitem{FREEDMAN1974354}
D.~Z. Freedman and E.~J. Weinberg, \emph{The energy-momentum tensor in scalar
  and gauge field theories},
  \href{https://doi.org/https://doi.org/10.1016/0003-4916(74)90040-2}{\emph{Annals
  of Physics} {\bfseries 87} (1974) 354}.

\bibitem{Ivanov:2019iec}
D.~Ivanov and S.~Liberati, \emph{{Testing Non-minimally Coupled BEC Dark Matter
  with Gravitational Waves}},
  \href{https://doi.org/10.1088/1475-7516/2020/07/065}{\emph{JCAP} {\bfseries
  07} (2020) 065} [\href{https://arxiv.org/abs/1909.02368}{{\ttfamily
  1909.02368}}].

\bibitem{Ji:2021rrn}
L.~Ji, \emph{{Wave Dark Matter Non-minimally Coupled to Gravity}},
  \href{https://arxiv.org/abs/2106.11971}{{\ttfamily 2106.11971}}.

\bibitem{Sankharva:2021spi}
K.~Sankharva and S.~Sethi, \emph{{Nonminimally coupled ultralight axions as
  cold dark matter}},
  \href{https://doi.org/10.1103/PhysRevD.105.103517}{\emph{Phys. Rev. D}
  {\bfseries 105} (2022) 103517}
  [\href{https://arxiv.org/abs/2110.04322}{{\ttfamily 2110.04322}}].

\bibitem{Barman:2021qds}
B.~Barman, N.~Bernal, A.~Das and R.~Roshan, \emph{{Non-minimally coupled vector
  boson dark matter}},
  \href{https://doi.org/10.1088/1475-7516/2022/01/047}{\emph{JCAP} {\bfseries
  01} (2022) 047} [\href{https://arxiv.org/abs/2108.13447}{{\ttfamily
  2108.13447}}].

\bibitem{Zhang:2023fhs}
H.-Y. Zhang and S.~Ling, \emph{{Phenomenology of wavelike vector dark matter
  nonminimally coupled to gravity}},
  \href{https://doi.org/10.1088/1475-7516/2023/07/055}{\emph{JCAP} {\bfseries
  07} (2023) 055} [\href{https://arxiv.org/abs/2305.03841}{{\ttfamily
  2305.03841}}].

\bibitem{Starobinsky:1979ty}
A.~A. Starobinsky, \emph{{Spectrum of relict gravitational radiation and the
  early state of the universe}}, {\emph{JETP Lett.} {\bfseries 30} (1979) 682}.

\bibitem{Turner:1987bw}
M.~S. Turner and L.~M. Widrow, \emph{{Inflation Produced, Large Scale Magnetic
  Fields}}, \href{https://doi.org/10.1103/PhysRevD.37.2743}{\emph{Phys. Rev. D}
  {\bfseries 37} (1988) 2743}.

\bibitem{Ford:1989me}
L.~H. Ford, \emph{{INFLATION DRIVEN BY A VECTOR FIELD}},
  \href{https://doi.org/10.1103/PhysRevD.40.967}{\emph{Phys. Rev. D} {\bfseries
  40} (1989) 967}.

\bibitem{Faraoni:2000wk}
V.~Faraoni, \emph{{Inflation and quintessence with nonminimal coupling}},
  \href{https://doi.org/10.1103/PhysRevD.62.023504}{\emph{Phys. Rev. D}
  {\bfseries 62} (2000) 023504}
  [\href{https://arxiv.org/abs/gr-qc/0002091}{{\ttfamily gr-qc/0002091}}].

\bibitem{Golovnev:2008cf}
A.~Golovnev, V.~Mukhanov and V.~Vanchurin, \emph{{Vector Inflation}},
  \href{https://doi.org/10.1088/1475-7516/2008/06/009}{\emph{JCAP} {\bfseries
  06} (2008) 009} [\href{https://arxiv.org/abs/0802.2068}{{\ttfamily
  0802.2068}}].

\bibitem{Golovnev:2008hv}
A.~Golovnev, V.~Mukhanov and V.~Vanchurin, \emph{{Gravitational waves in vector
  inflation}}, \href{https://doi.org/10.1088/1475-7516/2008/11/018}{\emph{JCAP}
  {\bfseries 11} (2008) 018} [\href{https://arxiv.org/abs/0810.4304}{{\ttfamily
  0810.4304}}].

\bibitem{Golovnev:2009ks}
A.~Golovnev and V.~Vanchurin, \emph{{Cosmological perturbations from vector
  inflation}}, \href{https://doi.org/10.1103/PhysRevD.79.103524}{\emph{Phys.
  Rev. D} {\bfseries 79} (2009) 103524}
  [\href{https://arxiv.org/abs/0903.2977}{{\ttfamily 0903.2977}}].

\bibitem{Golovnev:2009rm}
A.~Golovnev, \emph{{Linear perturbations in vector inflation and stability
  issues}}, \href{https://doi.org/10.1103/PhysRevD.81.023514}{\emph{Phys. Rev.
  D} {\bfseries 81} (2010) 023514}
  [\href{https://arxiv.org/abs/0910.0173}{{\ttfamily 0910.0173}}].

\bibitem{Moffat:2005si}
J.~W. Moffat, \emph{{Scalar-tensor-vector gravity theory}},
  \href{https://doi.org/10.1088/1475-7516/2006/03/004}{\emph{JCAP} {\bfseries
  03} (2006) 004} [\href{https://arxiv.org/abs/gr-qc/0506021}{{\ttfamily
  gr-qc/0506021}}].

\bibitem{Brownstein:2005zz}
J.~R. Brownstein and J.~W. Moffat, \emph{{Galaxy rotation curves without
  non-baryonic dark matter}},
  \href{https://doi.org/10.1086/498208}{\emph{Astrophys. J.} {\bfseries 636}
  (2006) 721} [\href{https://arxiv.org/abs/astro-ph/0506370}{{\ttfamily
  astro-ph/0506370}}].

\bibitem{Tasinato:2014eka}
G.~Tasinato, \emph{{Cosmic Acceleration from Abelian Symmetry Breaking}},
  \href{https://doi.org/10.1007/JHEP04(2014)067}{\emph{JHEP} {\bfseries 04}
  (2014) 067} [\href{https://arxiv.org/abs/1402.6450}{{\ttfamily 1402.6450}}].

\bibitem{Heisenberg:2014rta}
L.~Heisenberg, \emph{{Generalization of the Proca Action}},
  \href{https://doi.org/10.1088/1475-7516/2014/05/015}{\emph{JCAP} {\bfseries
  05} (2014) 015} [\href{https://arxiv.org/abs/1402.7026}{{\ttfamily
  1402.7026}}].

\bibitem{DeFelice:2016yws}
A.~De~Felice, L.~Heisenberg, R.~Kase, S.~Mukohyama, S.~Tsujikawa and Y.-l.
  Zhang, \emph{{Cosmology in generalized Proca theories}},
  \href{https://doi.org/10.1088/1475-7516/2016/06/048}{\emph{JCAP} {\bfseries
  06} (2016) 048} [\href{https://arxiv.org/abs/1603.05806}{{\ttfamily
  1603.05806}}].

\bibitem{deFelice:2017paw}
A.~de~Felice, L.~Heisenberg and S.~Tsujikawa, \emph{{Observational constraints
  on generalized Proca theories}},
  \href{https://doi.org/10.1103/PhysRevD.95.123540}{\emph{Phys. Rev. D}
  {\bfseries 95} (2017) 123540}
  [\href{https://arxiv.org/abs/1703.09573}{{\ttfamily 1703.09573}}].

\bibitem{Baumann:2022mni}
D.~Baumann, \emph{{Cosmology}}. Cambridge University Press, 7, 2022,
  \href{https://doi.org/10.1017/9781108937092}{10.1017/9781108937092}.

\bibitem{Salehian:2020bon}
B.~Salehian, M.~H. Namjoo and D.~I. Kaiser, \emph{{Effective theories for a
  nonrelativistic field in an expanding universe: Induced self-interaction,
  pressure, sound speed, and viscosity}},
  \href{https://doi.org/10.1007/JHEP07(2020)059}{\emph{JHEP} {\bfseries 07}
  (2020) 059} [\href{https://arxiv.org/abs/2005.05388}{{\ttfamily
  2005.05388}}].

\bibitem{Salehian:2021khb}
B.~Salehian, H.-Y. Zhang, M.~A. Amin, D.~I. Kaiser and M.~H. Namjoo,
  \emph{{Beyond Schr\"odinger-Poisson: nonrelativistic effective field theory
  for scalar dark matter}},
  \href{https://doi.org/10.1007/JHEP09(2021)050}{\emph{JHEP} {\bfseries 09}
  (2021) 050} [\href{https://arxiv.org/abs/2104.10128}{{\ttfamily
  2104.10128}}].

\bibitem{Irsic:2017yje}
V.~Ir\v{s}i\v{c}, M.~Viel, M.~G. Haehnelt, J.~S. Bolton and G.~D. Becker,
  \emph{{First constraints on fuzzy dark matter from Lyman-$\alpha$ forest data
  and hydrodynamical simulations}},
  \href{https://doi.org/10.1103/PhysRevLett.119.031302}{\emph{Phys. Rev. Lett.}
  {\bfseries 119} (2017) 031302}
  [\href{https://arxiv.org/abs/1703.04683}{{\ttfamily 1703.04683}}].

\bibitem{Bechtol:2022koa}
K.~Bechtol et~al., \emph{{Snowmass2021 Cosmic Frontier White Paper: Dark Matter
  Physics from Halo Measurements}},  in \emph{{Snowmass 2021}}, 3, 2022,
  \href{https://arxiv.org/abs/2203.07354}{{\ttfamily 2203.07354}}.

\bibitem{Schwartz:2014sze}
M.~D. Schwartz, \emph{{Quantum Field Theory and the Standard Model}}. Cambridge
  University Press, 3, 2014.

\bibitem{Himmetoglu:2009qi}
B.~Himmetoglu, C.~R. Contaldi and M.~Peloso, \emph{{Ghost instabilities of
  cosmological models with vector fields nonminimally coupled to the
  curvature}}, \href{https://doi.org/10.1103/PhysRevD.80.123530}{\emph{Phys.
  Rev. D} {\bfseries 80} (2009) 123530}
  [\href{https://arxiv.org/abs/0909.3524}{{\ttfamily 0909.3524}}].

\bibitem{Himmetoglu:2008zp}
B.~Himmetoglu, C.~R. Contaldi and M.~Peloso, \emph{{Instability of anisotropic
  cosmological solutions supported by vector fields}},
  \href{https://doi.org/10.1103/PhysRevLett.102.111301}{\emph{Phys. Rev. Lett.}
  {\bfseries 102} (2009) 111301}
  [\href{https://arxiv.org/abs/0809.2779}{{\ttfamily 0809.2779}}].

\bibitem{Himmetoglu:2008hx}
B.~Himmetoglu, C.~R. Contaldi and M.~Peloso, \emph{{Instability of the ACW
  model, and problems with massive vectors during inflation}},
  \href{https://doi.org/10.1103/PhysRevD.79.063517}{\emph{Phys. Rev. D}
  {\bfseries 79} (2009) 063517}
  [\href{https://arxiv.org/abs/0812.1231}{{\ttfamily 0812.1231}}].

\bibitem{Esposito-Farese:2009wbc}
G.~Esposito-Farese, C.~Pitrou and J.-P. Uzan, \emph{{Vector theories in
  cosmology}}, \href{https://doi.org/10.1103/PhysRevD.81.063519}{\emph{Phys.
  Rev. D} {\bfseries 81} (2010) 063519}
  [\href{https://arxiv.org/abs/0912.0481}{{\ttfamily 0912.0481}}].

\bibitem{Graham:2015rva}
P.~W. Graham, J.~Mardon and S.~Rajendran, \emph{{Vector Dark Matter from
  Inflationary Fluctuations}},
  \href{https://doi.org/10.1103/PhysRevD.93.103520}{\emph{Phys. Rev. D}
  {\bfseries 93} (2016) 103520}
  [\href{https://arxiv.org/abs/1504.02102}{{\ttfamily 1504.02102}}].

\bibitem{Mou:2022hqb}
Z.-G. Mou and H.-Y. Zhang, \emph{{A singularity problem for interacting massive
  vectors}}, \href{https://doi.org/10.1103/PhysRevLett.129.151101}{\emph{Phys.
  Rev. Lett.} {\bfseries 129} (2022) 151101}
  [\href{https://arxiv.org/abs/2204.11324}{{\ttfamily 2204.11324}}].

\bibitem{Clough:2022ygm}
K.~Clough, T.~Helfer, H.~Witek and E.~Berti, \emph{{Ghost Instabilities in
  Self-Interacting Vector Fields: The Problem with Proca Fields}},
  \href{https://doi.org/10.1103/PhysRevLett.129.151102}{\emph{Phys. Rev. Lett.}
  {\bfseries 129} (2022) 151102}
  [\href{https://arxiv.org/abs/2204.10868}{{\ttfamily 2204.10868}}].

\bibitem{Coates:2022qia}
A.~Coates and F.~M. Ramazano\u{g}lu, \emph{{Intrinsic Pathology of
  Self-Interacting Vector Fields}},
  \href{https://doi.org/10.1103/PhysRevLett.129.151103}{\emph{Phys. Rev. Lett.}
  {\bfseries 129} (2022) 151103}
  [\href{https://arxiv.org/abs/2205.07784}{{\ttfamily 2205.07784}}].

\bibitem{Capanelli:2024pzd}
C.~Capanelli, L.~Jenks, E.~W. Kolb and E.~McDonough, \emph{{Runaway
  Gravitational Production of Dark Photons}},
  \href{https://arxiv.org/abs/2403.15536}{{\ttfamily 2403.15536}}.

\bibitem{Mocz:2023adf}
P.~Mocz et~al., \emph{{Cosmological structure formation and soliton phase
  transition in fuzzy dark matter with axion self-interactions}},
  \href{https://doi.org/10.1093/mnras/stad694}{\emph{Mon. Not. Roy. Astron.
  Soc.} {\bfseries 521} (2023) 2608}
  [\href{https://arxiv.org/abs/2301.10266}{{\ttfamily 2301.10266}}].

\bibitem{Chen:2024pyr}
J.~Chen and H.-Y. Zhang, \emph{{Novel structures and collapse of solitons in
  nonminimally gravitating dark matter halos}},
  \href{https://doi.org/10.1088/1475-7516/2024/10/005}{\emph{JCAP} {\bfseries
  10} (2024) 005} [\href{https://arxiv.org/abs/2407.09265}{{\ttfamily
  2407.09265}}].

\bibitem{Chavanis:2011zi}
P.-H. Chavanis, \emph{{Mass-radius relation of Newtonian self-gravitating
  Bose-Einstein condensates with short-range interactions: I. Analytical
  results}}, \href{https://doi.org/10.1103/PhysRevD.84.043531}{\emph{Phys. Rev.
  D} {\bfseries 84} (2011) 043531}
  [\href{https://arxiv.org/abs/1103.2050}{{\ttfamily 1103.2050}}].

\bibitem{Chavanis:2011zm}
P.~H. Chavanis and L.~Delfini, \emph{{Mass-radius relation of Newtonian
  self-gravitating Bose-Einstein condensates with short-range interactions: II.
  Numerical results}},
  \href{https://doi.org/10.1103/PhysRevD.84.043532}{\emph{Phys. Rev. D}
  {\bfseries 84} (2011) 043532}
  [\href{https://arxiv.org/abs/1103.2054}{{\ttfamily 1103.2054}}].

\bibitem{Chavanis:2022fvh}
P.-H. Chavanis, \emph{{Maximum mass of relativistic self-gravitating
  Bose-Einstein condensates with repulsive or attractive
  |\ensuremath{\varphi}|4 self-interaction}},
  \href{https://doi.org/10.1103/PhysRevD.107.103503}{\emph{Phys. Rev. D}
  {\bfseries 107} (2023) 103503}
  [\href{https://arxiv.org/abs/2211.13237}{{\ttfamily 2211.13237}}].

\bibitem{Jain:2023ojg}
M.~Jain, M.~A. Amin, J.~Thomas and W.~Wanichwecharungruang, \emph{{Kinetic
  relaxation and Bose-star formation in multicomponent dark matter}},
  \href{https://doi.org/10.1103/PhysRevD.108.043535}{\emph{Phys. Rev. D}
  {\bfseries 108} (2023) 043535}
  [\href{https://arxiv.org/abs/2304.01985}{{\ttfamily 2304.01985}}].

\bibitem{Jain:2023tsr}
M.~Jain, W.~Wanichwecharungruang and J.~Thomas, \emph{{Kinetic relaxation and
  nucleation of Bose stars in self-interacting wave dark matter}},
  \href{https://doi.org/10.1103/PhysRevD.109.016002}{\emph{Phys. Rev. D}
  {\bfseries 109} (2024) 016002}
  [\href{https://arxiv.org/abs/2310.00058}{{\ttfamily 2310.00058}}].

\end{thebibliography}\endgroup
\end{document}